# Orion's Bar: Physical Conditions across the Definitive $H^+$ / $H^0$ / $H_2$ Interface

E.W. Pellegrini & J.A. Baldwin
*Physics and Astronomy Department, Michigan State University, 3270 Biomedical Physical Sciences Building, East Lansing, MI 48824*

G.J. Ferland
*Department of Physics and Astronomy, University of Kentucky, 177 Chemistry/Physics Building, Lexington, KY 40506*

Gargi Shaw
*Department of Astronomy and Astrophysics, Tata Institute of Fundamental Research, Mumbai-400-005, India*

S. Heathcote
*SOAR Telescope, Casilla 603, La Serena, Chile*

pelleg10@pa.msu.edu

**Abstract**

Previous work has shown the Orion Bar to be an interface between ionized and molecular gas, viewed roughly edge on, which is excited by the light from the Trapezium cluster. Much of the emission from any star-forming region will originate from such interfaces, so the Bar serves as a foundation test of any emission model. Here we combine X-ray, optical, IR and radio data sets to derive emission spectra along the transition from $H^+$ to $H^0$ to $H_2$ regions. We then reproduce the spectra of these layers with a simulation that simultaneously accounts for the detailed microphysics of the gas, the grains, and molecules, especially $H_2$ and CO. The magnetic field, observed to be the dominant pressure in another region of the Orion Nebula, is treated as a free parameter, along with the density of cosmic rays. Our model successfully accounts for the optical, IR and radio observations across the Bar by including a significant magnetic pressure and also heating by an excess density of cosmic rays, which we suggest is due to cosmic rays being trapped in the compressed magnetic field. In the Orion Bar, as we had previously found in M17, momentum carried by radiation and winds from the newly formed stars pushes back and compresses the surrounding gas. There is a rough balance between outward momentum in starlight and the total pressure in atomic and molecular gas surrounding the $H^+$ region. If the gas starts out with a weak magnetic field, the starlight from a newly formed cluster will push back the gas and compress the gas, magnetic field, and cosmic rays until magnetic pressure becomes an important factor.

## 1. Introduction

The interactions between light and winds from a newly-formed cluster and the molecular cloud in which the stars were born sculpts the geometry of the regions, produces the observed spectrum, and is a feedback mechanism that throttles the rate of star formation. To explore these processes in detail we are revisiting a series of well-studied nearby star forming regions. In particular we are examining objects with geometries viewed nearly edge-on, allowing us to measure the effect magnetic fields have at different depths as the starlight penetrates into the cloud. We use the observed stellar parameters, gas densities, and multi-wavelength emission-line spectrum to strongly constrain a numerical simulation of the physical conditions and emission along a ray from the central stars through the $H^+$, $H^0$, and $H_2$ regions. We include the effects of dust and of detailed molecule destruction and formation processes, and treat the detailed micro-physics of the $H^+$, $H^0$, and $H_2$ regions self-consistently. The only free parameters in simulations of a well observed cloud will be the cosmic ray density and magnetic field strength. In some cases the field can be directly measured. It is never possible to measure the cosmic rays directly so this approach is one of the few ways to infer their properties, although they are known to be produced in star-forming regions and can energize emission-line regions.

The modeling approach described in the preceding paragraph was recently applied to the Galactic H II region M17 (Pellegrini et al 2007; hereafter Paper I). That object contains a heavily obscured and nearly edge-on interface that is excited by about a dozen O stars. This interface is of particular interest because it is a rare case where the magnetic field can be measured in the adjacent $H^0$ region (or photodissociation region, the PDR). This is possible because radio continuum emission from the $H^+$ zone provides a background light source against which Zeeman polarization can be measured with the H I 21cm line (Brogan et al. 1999; Brogan & Troland 2001). The magnetic field is strong and magnetic pressure is important (see § 3 below). Combining the Zeeman measurements together with existing radio, infrared, and X-ray maps and new optical spectroscopy, we found that the structure of M17 is well described by a model in which the outward momentum carried by the stellar radiation field, together with pressure from a stellar wind-blown bubble, has compressed the gas and its associated magnetic field until the magnetic pressure built up sufficiently to be able to halt the process. The overall geometry is set by hydrostatic equilibrium. In addition, the density of cosmic rays is enhanced as a result of partial trapping of the charged cosmic ray particles by the compressed magnetic field, so that cosmic ray heating is important in atomic regions. We consider this to be a very natural cause-and-effect explanation of why the M17 gas cloud has taken on its present form.

Here we investigate whether this is also a good description of another edge-on interface – the well-known Bar in the Orion Nebula. Because it is very close to us (here we adopt a distance of 437 pc; Hirota et al. 2007), Orion is perhaps the best studied of all H II regions, with data across the entire electromagnetic spectrum. A schematic of the geometry is shown as Figure 8.4 of Osterbrock & Ferland (2006, hereafter AGN3). The ionizing radiation field is dominated by the hot O star $\theta^1$ Ori C. Light from this star is steadily dissociating the background molecular cloud, resulting in a blister type geometry in which the $H^+$ region is a hot skin on the surface of the molecular cloud. A large cavity has been carved out of the molecular gas, breaking out of the cloud on the side nearest the Earth so that we can see through the bubble to the $H^+$ region on the back wall (Zuckerman 1973; Balick et al. 1974; Baldwin et al. 1991, hereafter BFM91; Wen & O'Dell 1995; Ferland 2001; O'Dell 2001). The hot gas filling this cavity has recently been detected by Güdel et al (2008).

Given the important role that magnetic pressure plays in M17, it is natural to ask whether it might also be important in Orion. While there are no direct observations of the field strength in the atomic gas associated with the Orion Bar, the Orion complex shows a well-structured polarization pattern that drops to a low level of polarization in the Bar, suggesting that the magnetic field in the Bar is directed more or less along our line of sight (Schleuning 1998). This is the expected orientation for an initially tangled magnetic field frozen into gas that has been compressed by radiation pressure or stellar winds from $\theta^1$ Ori C. There are Zeeman measurements for the Orion Veil (a foreground structure that is associated with the Orion Molecular Cloud) showing that $B_{los}$ ~ 50 μG (Brogan et al. 2005). This is already an order of magnitude greater than the Galactic background of 5-10μG (Tielens & Hollenbach 1985), even though $B_{los}$ only represents the line of sight component. Detailed analysis suggests that the ratio of magnetic to gas pressure in the Veil is large (Abel et al. 2004, 2006). A further indication of the presence of a strong *B* field in the Veil is an indirect study using infrared line ratios (Abel and Ferland 2006), which indicated that along the line of sight to $\theta^1$ Ori C the ratio of magnetic/gas pressure is $P_{mag} / P_{gas} > 1$. Aside from the suggestive results of the Veil, turbulent velocities in the atomic region of the background cloud are supersonic and it has been suggested that magnetic fields are responsible (Kristensen et al. 2007; Roshi 2007). These results lend plausibility to the possible presence of a strong magnetic field in the Bar region.

The Bar appears as a bright ripple on the background ionization front, lying at a projected distance from $\theta^1$Ori C of 111 arcsec (0.23 pc). In an important series of papers Tielens et al (1993), Tauber et al. (1994), and Young Owl et al. (2000) showed that the $H^0$ and $H_2$ regions in the Bar are easily resolved on the sky as separate structures displaced from each other in a way that clearly demonstrates that the Bar is indeed a roughly edge-on interface. They showed that most of the observed emission in PAHs, [O I], [C II], $H_2$ and the $^{12}$CO J=1–0 lines can be understood as coming from a homogeneous region with density $n_H \sim 5\times10^4$ cm$^{-3}$. However, they argued that large (9 arcsec = 0.02 pc) clumps with about 20 times higher density must be imbedded in this homogeneous medium to produce the observed high-level CO lines (J=14–13, 7–6) and also the $HCO^+$ and HCN emission. Indeed, their interferometer images directly show clumpy structures in these lines but not in the many lines attributed to the homogeneous medium.

Those investigations considered only the molecular and neutral atomic regions of the Bar. The density chosen by Tauber et al. (1994) for their model was inferred from the observed offsets of the $H_2$ and $^{12}$CO emission in the Bar relative to the ionization front, combined with the column density required to allow the observed CO emission to be produced. In doing so, the authors assumed constant gas density in the $H^0$ region, arguing that turbulence dominated the pressure.

Van der Werf et al. (1996) combined new $H_2$ observations with the existing data set of the PDR and postulated that $H_2$ emission from the interclump medium required a filling factor less than unity for the interclump gas. Contradicting this result, Allers et al. (2005) also modeled the interclump region with constant pressure, not constant density. They concluded that a filling factor of unity described the region well. However they also found that an extra heating agent must be present in the Bar, but were unable to establish what it might be.

Here we investigate the physical conditions across the Bar that are implied by the variations of its spectrum across its full width. Like the previous work, we use the observed gas density, stellar parameters, and emission peak offsets. However, we consider the emission from the $H^+$ and PDR regions together in a self-consistent picture. By more fully understanding what determines the structure of the Bar we can better understand the general nature of such interfaces. Here we focus on a line of sight through the interclump medium, not on the clumps described by Young Owl et al. (2000) and others. Like many (but not all) previous papers we conclude that the interclump medium dominates and determines the overall structure of the Bar. We find that the observed densities and the radial extent of the various stratified emission regions through the Bar are in fact the natural consequences of a cloud in roughly hydrostatic equilibrium, in which magnetic pressure and cosmic ray heating play major roles in the $H^0$ region. This study, combined with the companion work by Shaw et al. (2008; hereafter Paper III), which focuses on the detailed $H_2$ emission spectrum, lead to a better picture of the nature of these interfaces.

## 2. The Observational Data Set

The Orion Nebula has been observed extensively at all wavelengths. For this study of the Bar we draw on the following published data sets. From Tauber et al. (1994) we use maps of the $^{12}$CO and $^{13}$CO emission lines, and their summary of previous mid-IR observations of the CO 7–6, CO 14–13, O I λ63 μm, O I λ143 μm, C II λ158 μm, Si II λ35 μm, C I λ609 μm and C I λ370 μm emission lines. The $H_2$ 1-0 S(1) data are originally from van der Werf et al. (1996), who find an average surface brightness of $5.9\times10^{-15}$ erg $s^{-1}$ $cm^{-2}$ $arcsec^{-2}$. However, the $H_2$ surface brightness varies significantly along the Bar. A05 and Young Owl et al. (2000) used the van der Werf et al. $H_2$ data set but extracted surface brightness profiles along different narrow lines perpendicular to the Bar. A05 did this for a particularly bright region and found a peak brightness of about $9\times10^{-15}$ erg $s^{-1}$ $cm^{-2}$ $arcsec^{-2}$. Young Owl et al. (2000) did the same for a fainter region which is along the same cut for which most of the other published molecular data used here were measured, and found the peak $S(H_2)$ 1-0 S(1) = $3.3\times10^{-15}$ erg $s^{-1}$ $cm^{-2}$ $arcsec^{-2}$. For consistency we adopt the Young Owl et al. $H_2$ surface brightness profile. The peak surface brightnesses of these emission lines are summarized in Table 1 with observed values in column 2 and results of our three models described below in columns 3 through 5.

Besides $H_2$ and CO, many other molecules are detected in the Bar. Here we will also compare our predictions with observations for $CO^+$ (Störzer et al. 1995), $SO^+$ (Fuente et al. 2003), CN (Simon et al. 1997), CS (Simon et al. 1997, Hogerheijde et al. 1995), SiO (Schilke et al . 2001), and SO (Leurini et al. 2006).

Pogge et al. (1992) made optical passband Imaging Fabry-Perot maps of Hα, Hβ, [O III] λ5007, [N II] λλ6548, 6583, [S II] λλ6716, 6731 and He I λ6678, and in their Figure 5 show a density profile across the Bar, measured from the [S II] λ6716/λ6731 intensity ratio. Wen & O'Dell (1995) used the Pogge et al. data to make a 3-dimensional map of the $H^+$ region in Orion and their Figure 4 shows intensity profiles across the Bar for several lines including Hα. García-Díaz & Henney (2007) mapped a number of optical emission lines by taking an extensive grid of echelle long-slit spectra, and in their Figure 9 show a density profile across the Bar (from the [S II] intensity ratio) that is very similar to the Pogge et al. (1992) result. From these papers, we have especially depended on several figures showing line intensities along the cuts across the Bar shown here in Figure 1.

To this existing data set we added an intensity profile across the Bar in [S II] λ6713+λ6731, measured from a continuum-subtracted, narrow-band [S II] image taken on 06 April, 2007 with the Southern Astrophysical Research (SOAR) Telescope[1]. The [S II] filter is 45Å wide and is centered at λ6723, while the continuum filter is centered at 6850Å with a width of 95Å.[2] The SOAR Optical Imager (SOI) was used; it provides a 5×5 arcmin field of view with a 2×2 binned pixel scale of 0.15 arcsec/pixel and yielded images with 0.6 arcsec FWHM in both filters. The images were processed in IRAF in the usual way, then the continuum image was scaled to unsaturated stars in the nebula and subtracted from the emission line image. Next, the continuum-subtracted [S II] image was flux calibrated using the emission-line surface brightness measurements from the long-slit spectrophotometry of BFM91, following the example of O'Dell and Doi (1999). Finally, the [S II] intensity profile was measured along the same cut as is shown in Figure 1 for the Wen & O'Dell Hα profile.

Figure 2 combines together the four key intensity profiles that we will use to constrain our models of the Bar. The ionizing radiation comes in from the left. Using the new SOAR [S II] images, the different profiles have all had their zero points set to match the distance from the peak in the [S II] emission along each cut. There are several key features to note. The ionization front (IF) is assumed to be at the position of the peak [S II] emission. Hα emission comes from a broad plateau before (to the left of) the IF. The $H_2$ brightness profile peaks about 12 arcsec after the IF

[1] The Southern Astrophysical Research Telescope is a joint project of Michigan State University, Ministério da Ciência e Tecnologia-Brazil, the University of North Carolina at Chapel Hill, and the National Optical Astronomy Observatory. Further information about SOAR and its instruments may be found at www.soartelescope.org.

[2] We are grateful to Dr. Frank Winkler for loaning us the filters.

(i.e. deeper into the cloud). In the $^{12}$CO J = 1–0 line the peak intensity occurs about 20 arcsec after the IF and is very broad.

As can be seen in Figure 1, these cuts across the Bar are not at identical locations. The Wen & O'Dell H$\alpha$ profile and our [S II] profile are averaged over 20 arcsec wide swathes extending in PA 322 deg from $\theta^1$ Ori C, while the $^{12}$CO (from Tauber et al. 1994) and $H_2$ profiles (Young Owl et al. 2000) are taken in PA 315 deg crossing the Bar about 60 arcsec SE of the H$\alpha$ and [S II] cuts. However, close examination of maps of the [S II] ratio in the vicinity of the Bar (e.g. Fig. 11 of Henney et al. 2005a), and the similarities of the same lines published for a different position along the Bar (Young Owl et al. 2000) indicates that the relative offsets and shapes of the different line profiles are typical of the interclump component of the Bar as a whole, and therefore are valid criteria for fitting our models.

## 3. A ray through the $H^+$ / $H^0$ / $H_2$ Layers of the Bar

### 3.1 Numerical simulations of the Bar

The bright ridge of emission seen as the Bar in Figure 1 is a region where the gas goes from $H^+$, closest to the central stars, through $H^0$ when ionizing radiation has been attenuated, eventually becoming $H_2$ when ultraviolet light is sufficiently extinguished (see also Figures 8.4 and 8.6 of AGN3). We derive physical conditions across the Bar by comparing predicted and observed spectra at various distances away from the star cluster (as represented by $\theta^1$ Ori C). The simulations are done with the spectral synthesis code Cloudy. Last described by Ferland et al. (1998),[3] Cloudy has since been updated to include a large model of the hydrogen molecule (Shaw et al. 2005), more complete grain physics (van Hoof et al. 2004) and the chemistry of a PDR (Abel et al. 2005). We used the publicly available version of Cloudy, but with updated $H_2$ collision rates as described in Paper III.

We begin the calculation at the ionized or illuminated face of the $H^+$ region and follow a beam of starlight away from the central cluster into the molecular cloud. The $H^+$, $H^0$ and $H_2$ regions are actually a flow from cold molecular gas through the atomic region into hot and ionized gas with the ionizing radiation gradually eating into the molecular cloud. However, we approximate the situation here with a hydrostatic model. We will present predicted and observed quantities along this line from the stars into the molecular cloud, with the ionizing radiation coming in from the left in all relevant figures.

The most important assumption in all of this is that the conditions within the various regions are related to one another by continuous variations in the gas, radiation, and magnetic pressures. The equation of state, the relationship between these pressures and the density, is described in Paper I and further in Appendix A. Because the structure of the $H^+$, $H^0$ and $H_2$ regions are all the result of a single calculation, with a single set of initial conditions, we minimize the number of adjustable parameters compared to the number of observational constraints. As will be shown in section 4.1 our best model will have only 6 arbitrary input parameters, but will satisfactorily match 17 observational constraints. For example, the UV radiation that penetrates into the PDR is the result of a detailed treatment of radiation transport through the $H^+$ region. By treating the $H^+$ region and PDR together in a single calculation, the radiation affecting the PDR is guaranteed to be consistent with the observed properties of $\theta^1$ Ori C and the gas density of the $H^+$ region. The emission from ionized and neutral atoms, grains and PAH chains, and several molecular species including $H_2$ and CO, are all predicted by self-consistently treating the microphysics of the gas. Starlight is attenuated by each layer and passed on to more distant regions. All models were also required to reproduce the observed 111 arcsec projected offset of the [S II] emission from $\theta^1$ Ori C, which clearly defines the Bar's ionization front.

[3] Many additional physical processes described in Cloudy are documented and referenced in Hazy, the approximately 1500 page Cloudy manual. It can be downloaded from ftp://gradj.pa.uky.edu/gary/cloudy_gold/docs

The location of the illuminated face of the $H^+$ region, and the gas density there, are set by the H$\alpha$ emission profile. There is a steep drop in the H$\alpha$ surface brightness where the $H^+$ zone ends and the $H^0$ zone begins. There is no similar sharp rise to mark the illuminated face of the Bar because the light emitted by the edge-on Bar structure is diluted by additional emission coming from parts of the $H^+$ region which are on the background cavity surface but which lie along the same line of sight. Therefore, the initial density and radius of our models can only be constrained by their effects beyond the illuminated face. These initial parameters are chosen so that the observed H$\alpha$ and [S II] surface brightness and [S II] doublet ratio are matched. Since we assume hydrostatic equilibrium the gas pressure increases as starlight with its associated momentum is absorbed. This causes an increase in density and emission measure, $n^2\, dV$, producing the observed brightness profile.

The parameters assumed in our simulations are largely observationally based and are summarized next. The main free parameters will be the magnetic field intensity and cosmic ray density. We assume flux freezing and a scrambled magnetic field to relate the gas density and field.

Three models are presented here, designated (1) Gas Pressure Model, (2) Magnetic Field Model and (3) Enhanced Cosmic Ray Model. They all share the following assumptions:

- The grain optical properties, described by BFM91, are based on the observed extinction in Orion. Many previous investigations, going as far back as Tielens & Hollenbach (1985), assume standard ISM extinction. Orion grains have a more nearly grey dependence of extinction on wavelength compared with ISM grains, changing the structure of the layers. The Orion size distribution is deficient in small particles so produces less heating of the gas by grain electron photo-ejection.
- The PAH abundance is $n_{PAH}/n_{H^o} = 3\times10^{-7}$ (Draine et al. 2007) with a power-law distribution of PAH sizes with 10 size bins, according to Bakes & Tielens (1994).
- We assume overall hydrostatic equilibrium as described in Appendix A, with magnetic fields (when included) described by Eq. A1 with $\gamma = 2$.
- The gas-phase abundances are the standard Cloudy values for H II regions, and are largely based on observations of the Orion Nebula described by BFM91, Osterbrock et al (1992), and Rubin et al (1991, 1993). These abundances are listed in Table 2 of Paper III.
- We have taken the distance to the Orion Nebula to be 437±19 pc (Hirota et al. 2007). This is a VLBI parallax measurement, which should be more accurate than previous results.
- All models are constrained to reproduce the observed [S II] ratio I($\lambda$6716)/I($\lambda$6731) = 0.63 ± 0.05, converted from the reported electron density of $n_e \sim 3200$ cm$^{-3}$ at the peak of the Bar (Pogee et al. 1992). This is consistent with the [S II] ratios reported by García-Díaz & Henney (2007) for the Bar. The calculations predict the full spectrum including this ratio. We match this ratio instead of using a deduced density since the conversion from ratio to density depends on the kinetic temperature and ratio of electrons to atoms.
- The observed stellar X-ray emission caused by wind activity immediately around the Trapezium stars is modeled with a bremsstrahlung distribution with a temperature $T = 10^6$K and an integrated luminosity $L_x = 10^{32.6}$ erg s$^{-1}$ over the 0.5–8 keV passband (Feigelson et al. 2005). This represents the X-ray emission from just $\theta^1$ Ori C. There are additional X-ray point sources in this region besides $\theta^1$ Ori C, but their positions along the line of sight are poorly known. Omitting them is not likely to be a large source of error because $\theta^1$ Ori C accounts for 68% of the total observed X-ray flux from the Orion region. The diffuse X-ray emission detected by Güdel et al (2008) is much weaker than the stellar contribution.

- A constant turbulent velocity was assumed. The observed $^{13}CO$ line width is 1.8 km $s^{-1}$ FWHM (Tauber et al. 1994), while the $H_2$ line widths are in the range 2–4 km $s^{-1}$ after allowance for a poorly known instrumental profile (A05). Here we adopt a constant turbulent velocity of 2 km $s^{-1}$ FWHM throughout the nebula and include it as a source of pressure in our models.

The three models are described below. Each builds on the results from the preceding one by adding additional physical processes.

**3.2 The Gas Pressure Model**

This first model represents the BFM91 hydrostatic model of the $H^+$ region. It does not include a magnetic field but does include radiation and turbulent pressure. We shall refer to this as the gas pressure model even though turbulent and radiation pressures make significant contributions. Cosmic ray heating and ionization are included in the calculation using a cosmic ray density set to the Galactic average of $2.6\times10^{-9}$ $cm^{-3}$ (Williams et al. 1998). The ionization rate per particle corresponding to this density is $2.5\times10^{-17}$ $s^{-1}$ for $H^0$ and $5\times10^{-17}$ $s^{-1}$ for molecular $H_2$. For this model the heating due to cosmic rays is mostly insignificant, with the cosmic rays peaking at 15% of the heating in the coldest and most neutral regions, where the starlight is heavily extinguished. Collisional heating of grains (Drain 1978) is the most important heating mechanism in the molecular region, providing 70% of the total heating. In this model we find the heating in the deep molecular gas is not dominated by cosmic rays but by grain heating processes, consistent with the conclusions of A05.

Figure 3 shows our assumed geometry. The Bar is a thin slab tilted slightly to our line of sight. Because the hydrogen ionization front is thin, the observed [S II] profile on the sky is very sensitive to the angle of inclination. The thin slab extends 0.115 pc along the line of sight and is inclined 7 deg relative to the line of sight. This is very similar to many previous models of the Bar's geometry (eg. Tielens et al. 1993; Hogerheijde et al. 1995; Wen & O'Dell 1995; Walmsley et al. 2000; A05).

For a given incident ionizing flux $\Phi(H)$ photons $s^{-1}$ $cm^{-2}$, the shape and intensity of the [S II] profile is set by the initial hydrogen density $n_0(H)$. This also defines the position of the hydrogen ionization front. The problem is that we do not initially know either $\Phi(H)$ or $n_0(H)$.

The number of ionizing photons per second $Q(H)$ emitted by $\theta^1$ Ori C and hence the ionizing flux $\Phi(H)$ incident upon the slab is uncertain. Although the star's spectral type is clearly O6.5, there is a significant range in the $Q(H)$ and effective temperature values that are thought to be appropriate for even a "normal" O6.5 star. At the high end is $Q(H) = 10^{49.23}$ $s^{-1}$ with $T_{eff}$ = 42,300 K (Vacca et al. 1996), while Hanson et al. (1997) adopted the much lower value $Q(H) = 10^{48.89}$ $s^{-1}$ and $T_{eff}$ = 41,200 K for this spectral type. Intermediate values are suggested by Smith et al. (2002) and Sternberg et al. (2003). In addition to this uncertainty, $\theta^1$ Ori C has an unusually strong magnetic field, which is thought to channel the flow of stellar winds from the star in ways that modulate the observed spectrum (Wade et al. 2006). This could produce a significantly anisotropic radiation field.

Given the above uncertainty, we explored a range of values of $T_{eff}$ and $Q(H)$. Our models have only a very slight dependence on the exact value of $T_{eff}$, but the value used for $Q(H)$ clearly does matter. Changing $Q(H)$ while matching all the observations of the $H^+$ region, including the density $n_e$ which is a measured quantity, does not significantly affect the predictions for the $H^0$ and $H_2$ regions, but it does change the deduced distance between $\theta^1$ Ori C and the ionization front. Since the projection of this distance on the sky is fixed, the position of the slab along our line of sight must change relative to $\theta^1$ Ori C to match the observed geometry. For the largest $Q$(H) value cited above the slab would have to lie about 0.125 pc farther away from us than in the adopted model, while maintaining the same inclination angle relative to our line of sight. We eventually adopted the Kurucz (1979) stellar model as the continuum shape of $\theta^1$ Ori C with $Q$(H) = $10^{49.00}$ $s^{-1}$ and $T_{eff}$ = 39,700 K. This $Q$(H) is the same as was used in the 3D model for which Wen and O'Dell (1995) show figures, and also was used in the hydrodynamic wind models computed by Henney et al. (2005a), simplifying comparison to those papers.

However, this still left many combinations of the Bar's thickness $l$ along the ray from $\theta^1$ Ori C and in its density $n_0$(H) at the illuminated face, and consequently of the radial gas density distribution $n_H(r)$, which were consistent with the constraints used so far. We used the additional constraints of the Hα and [S II] brightness profiles together with the [S II] λ6716/λ6731 intensity ratio to determine the remaining properties of the entire $H^+$ zone. The electron density $n_e \sim n_H$ is directly measured in the region where the [S II] lines are formed, so the [S II] surface brightness was used to determine the distance $h$ through the Bar *along the line of sight* using

$$S([\mathrm{S\ II}]) \propto n_{S^+} n_e h. \qquad (1)$$

Then the Hα surface brightness at other points on the Bar was used to determine the relationship between density and position on the sky for the known $h$. Besides increasing the peak surface brightness, a higher $n_H$ also decreases the deduced thickness $l$ of the $H^+$ region.

The result of this was a model, with predicted emission line strengths and magnetic field summarized in column 3 of Table 1, which reproduced the structure and emission of the $H^+$ region. In this model, $h$ = 0.115 pc, the illuminated face of the cloud lies 0.114 pc from $\theta^1$ Ori C, $l$ = 0.141 pc, and $\Phi$(H) = $6.45\times10^{12}$ $s^{-1}$ $cm^{-2}$.

The bottom row of Figure 4 shows the pressure sources and $H^+$, $H^0$ and $H_2$ density distributions calculated for the gas pressure model. This figure also includes the same plots for the other two models which are described in the following two sections.

Figure 5 shows, again for all three models, how well the computed surface brightness distributions of [S II], Hα, $H_2$ and CO lines match the observations. The results for the gas pressure model are shown in the left-hand column. We computed the conditions and locally emitted spectrum for each point along a ray from the central star through the layers shown in Figure 3. For the case of optically thin lines ([S II], Hα and $H_2$), the comparisons of the models to the observed surface-brightness distributions were made by integrating the volume emissivity along the line of sight into the cloud according to

$$S_i = \int_0^h \frac{\varepsilon_i(\vec{r})}{4\pi} 10^{-dA_i(\vec{r})/2.5}\, dh \qquad (2)$$

The integration is along the line of sight into the modeled region of thickness $h$, $\varepsilon_i$ is the volume emissivity for the $i^{th}$ line, $\vec{r}$ is the radial vector from $\theta^1$ Ori C, and $dA_i$ is the amount of internal reddening, in magnitudes, providing a correction to the observed line intensities for internal extinction by dust. The computed surface brightness of the visible-passband emission lines shown in Figure 5 and listed below in Table 1 have been increased by a factor 1.5 to account for an additional component reflected from dust in the molecular cloud (Wen and O'Dell 1995; O'Dell et al. 1992). At longer wavelengths the Orion dust does not scatter efficiently (BFM91) so no similar correction is needed for the infrared and mm-wavelength lines.

The reddening correction is specific for each line according to the $R$ = 5.5 reddening curve used for Orion. The mid-IR and longer wavelength lines are unaffected by extinction due to their long wavelength and the $H^+$ region observations have been dereddened (BFM91, Wen & O'Dell 1995). Only the analysis of the $H_2$ 2.121μm line is affected by internal extinction. This amounts to a flux decrement of roughly 40% for the models presented below. Increasing the depth of the slab along the line of sight (*i.e.* parallel to the ionization front) increases the intensities of all lines except that of the $H_2$ 2.121μm line which stays approximately constant due to the effects of extinction by dust.

The $^{12}$CO J = 1–0 line is also different because of its large optical depth. For optically thick thermalized lines the emission is directly related to the kinetic temperature of the gas via the antenna temperature $T_{antenna}$. Deep in the cloud the density is very high and the optical depth increases rapidly. The antenna temperature quickly approaches the kinetic temperature according to the equation

$$T_{antenna} = T_{kin}(1-e^{-\tau}) \qquad (3),$$

where $\tau$ is the CO line optical depth along our line of sight into the cloud and $T_{kin}$ is the computed spin temperate of the $^{12}$CO levels. Our calculations solve for τ toward the illuminated face, which we then scale to account for the enhanced path-length caused by our viewing angle (see Figure 3). The computed CO surface brightness curves in Fig. 5 show $T_{antenna}$ starting at the edge of the slab farthest from us, which from our viewing angle is projected to lie closest to $\theta^1$ Ori C.

The top-left panel in Figure 5 shows the excellent fit of the gas pressure model's [S II] surface brightness distribution to the data. The gas pressure model's match to the observed Hα surface brightness profile (Figure 5, second panel down in left-hand column), as well as [O III] 5007Å and [N II], with no further adjustments to the model, validates our assumptions.

The gas pressure model accurately describes the $H^+$ region, but does not correctly predict the positions of the $H_2$ and $^{12}$CO emission peaks on the sky, as can be seen in the left-hand panels of Figure 5. This model was tuned to reproduce the observed $H^+$ emission region using only gas, radiation, and turbulent pressures. In this situation, the gas density in the $H^0$ region that is required to maintain hydrostatic equilibrium is $1.4\times10^5$ cm$^{-3}$ (Fig. 4). Since the depth of the $H^0$ region is set by the path length required to reach an $A_v \sim 1$ where $H_2$ forms, this high-density $H^0$ region is quite narrow. The result of the high density is that the computed $H_2$ emission peak occurs two times closer to the ionization front than is observed (5 arcsec predicted separation rather than the observed 12 arcsec), as is seen in Fig. 5. This result is consistent with the theory of $H_2$ emission as presented by Black and van Dishoeck (1987) and Draine and Bertoldi (1996). For further discussion of $H_2$ we refer the reader to Paper III. For the same reason, the CO emission also peaks at a point too close to $\theta^1$ Ori C. Table 1 shows that the predicted peak $H_2$ emission is equal to 1.51 times the peak value, while the predicted CO emission is several times fainter than is observed.

### 3.3 The Magnetic Pressure Model

This second model uses the same constraints as the gas pressure model described above, but includes a magnetic field similar in strength to that seen in the Veil (Abel et al. 2005). The cosmic ray density was maintained at the Galactic background level. We ran a series of models with increasing values of the strength of the magnetic field at the illuminated cloud face, in the same way that we had done in Paper I for M17. The initial magnetic field sets the field strength throughout the model since we assume flux freezing (Eq. A1 in the Appendix). The resulting magnetic pressure contributes to the total pressure according to Eq. A2. The center two panels in Figure 4 show the pressure contributions and densities as a function of depth in the final version of this model.

In the $H^+$ region where the temperatures are of the order $10^4$ K, gas pressure still dominates, so the gas pressure and magnetic pressure models are very similar in this region. The observed [S II] density and [S II] and Hα surface brightness profiles are matched by either model using the same initial conditions.

In the $H^0$ region the temperature drops and the gas density increases. According to equation A1, the magnetic field is amplified as well, so magnetic pressure support becomes important and further compression of the gas is halted. With a lower average density in the $H^0$ region a longer path length is required to absorb the UV photons that prevent $H_2$ from forming, so the $H^0$ region becomes more extended. Stronger fields at the illuminated face produce more magnetic pressure in the $H^0$ region, resulting in a lower density and larger thickness. We changed the initial magnetic field at the face of the cloud so that the $H_2$ emission peak occurred at its observed offset of about 10 arcsec (0.021 pc) from the ionization front. We found that an initial field of 8 μG best fits the observed brightness distribution of the $H_2$ emission line. In the $H^0$ region this field is $\langle B \rangle$ = 438 μG and the density in the $H^0$ region is $8\times10^4$ $cm^{-3}$, about 2 times lower than the density in the gas pressure model.

With the magnetic pressure model, the brightness and position of the $H_2$ peak now match the observations (Fig. 5). However, the model still does not reproduce the position and antenna temperature of the $^{12}$CO J = 1–0 emission peak or the high surface brightness in the higher-level $^{12}$CO lines (Table 1).

**3.4 The Enhanced Cosmic Ray Model**

The magnetic field model fails because it does not provide sufficient heating in the deeper parts of the $H^0$ region. Based on the results from M17 in Paper I, we next explored the effect of assuming that cosmic ray particles are trapped by the compressed magnetic field, so that the cosmic ray density is also increased. The increased density of cosmic rays acts as an additional heat source, which becomes important in the region emitting the $H_2$ and $^{12}$CO lines. The cosmic rays also increase the ionization level in molecular regions, increasing the speed of ion–molecule interactions and enhancing the CO formation rate. This moved the $^{12}$CO J = 1–0 peak inwards toward the central stars. A second effect is that the kinetic temperature is increased and the $H_2$ emission is enhanced deep into the cloud. The model with enhanced cosmic rays produces extended $H_2$ emission beyond the emission peak. We ran models with the cosmic ray density increased by different values over the Galactic background density, up to the point where the cosmic ray energy density is in equipartition with the magnetic energy density. Figure 6 shows the dependence of the computed $^{12}$CO J=1–0 brightness temperature on the cosmic ray density. The observed CO brightness temperature, peak $H_2$ intensity, and shapes and positions of the $H_2$ and CO profiles all are matched best by a cosmic ray density in equipartition with the magnetic field. That is the cosmic ray density we adopted in our final best model, which we call the enhanced cosmic ray model.

The average magnetic field, weighted by the 21 cm opacity $T_{spin}/n(H^0)$, for the enhanced cosmic ray model is $\langle B \rangle$ = 516μG, almost 100μG higher than in the magnetic field model, where the cosmic ray density was set to the Galactic background. The difference is due to the weighting of $B$ by $T_{spin}/n(H^0)$ that is used to calculate $\langle B \rangle$. The ratio is the 21 cm opacity used by Zeeman measurements to derive $B$. In the enhanced cosmic ray model, deep regions of the cloud which would normally be fully molecular have a significant amount of $H^0$ produced by cosmic ray dissociation. The effects of the cosmic rays on the chemistry are shown in Figure 4. Neutral hydrogen persists deep into the molecular core, where $B$ peaks at about 530μG. Here the cosmic ray density is enhanced by a factor of $10^{3.6}$ over the Galactic background. This corresponds to an ionization rate of $1\times10^{-13}$ $s^{-1}$ for $H^0$ and $2\times10^{-13}$ $s^{-1}$ for $H_2$. In the case of the magnetic model, the neutral hydrogen does not coexist with $H_2$. This extended distribution of neutral hydrogen is a result of the enhanced cosmic rays and is not dominated by FUV photons. It would be incorrect to consider it part of the classically defined "PDR", but it will still affect the weighted $<B>$ an observer will measure.

## 4. Discussion

### 4.1 The parameters needed to fit the observations

Our final model (the enhanced cosmic ray model) provides an integrated description of the ionized, neutral and molecular regions of the Bar. It is based on the idea that the pressure of photons from $\theta^1$ Ori C has compressed the surface of the molecular cloud, and along with it a magnetic field that was already present, until the combination of gas, magnetic and turbulent pressure became high enough to halt the compression. This model indicates that the Bar is in (quasi)hydrostatic equilibrium. This picture of the Orion Bar, combined with straightforward assumptions about the geometry, provides a good fit to the observed parameters. It reproduces the observed surface brightness profiles in the sense of both the position and the peak brightness of the H$\alpha$, [S II], $H_2$ and $^{12}CO$ emission lines. As is shown in Table 1, it satisfactorily reproduces the [S II] $\lambda6716/\lambda6731$ ratio, the integrated strengths of many other lines from the $H^+$ region, and the peak surface brightness of important atomic lines from the $H^0$ region including [O I] $\lambda63\mu$m, [Si II] $\lambda34.1\mu$m and [C II] $\lambda158\mu$m. We do not have a good optical spectrum with the slit set across the Bar, but we checked the predicted optical emission lines in Table 2 and verified that the computed optical spectrum is similar to that found by BFM91 at their position 5 on the west side of the Orion Nebula and that no unusual lines are predicted to be strong. Position 5 represents a region with a density comparable to the Bar. Our final model provides a quite good fit to the data. We matched 17 observed properties with 6 free parameters: the location of the illuminated face and the gas density at the face, the tilt of the IF and its depth along the line of sight, the magnetic field strength, and the cosmic ray density.

We arrived at this model in three steps. The properties of the $H^+$ region were determined by fitting a hydrostatic model that included only gas, radiation, and turbulent pressure – the gas pressure model. The location of the illuminated face and the gas density at the face, the tilt of the IF and the depth of the IF along the line of sight needed to be adjusted to reproduce the properties of the $H^+$ region. However, the resulting gas pressure model produced an $H^0$ region that is too narrow. To accurately fit the distance from the IF to the $H_2$ emission peak, it was necessary to add magnetic pressure. For the case of hydrostatic equilibrium this results in a decrease in the gas density, causing the $H^0$ region to become more extended and pushing the peak emission of $H_2$ and $^{12}CO$ to the observed value. However, this magnetic pressure model still failed to reproduce the observed $H_2$ and CO emission from deep in the cloud because the kinetic temperature was too low. We propose that the extra heating comes from cosmic rays trapped by the compressed magnetic field, the same thing that appears to be happening in M17. This fully reproduces the emission and geometry.

### 4.2 Predicted column densities of additional molecules

Cloudy includes 94 molecules in its calculations, using data mainly from the UMIST data base[4] (Abel et al. 2005). Chemical fractionation is not included in UMIST, so we cannot now deal with molecular isotopes. The UMIST data base does not include information about the internal structure of molecules. Therefore, we can only compute the molecular space density and derive column densities for this full set of molecules. Here we compare results for the subset of molecules which we consider most reliable and for which observations are available. Figure 7 shows this comparison as a function of projected distance from the IF for $CO^+$, CN, $SO^+$, SO, CS and SiO. Our models directly predict the volume density of each molecule. We have then computed a predicted column density $N_j(r)$ at an angular offset $r$ arcsec from the IF using

$$N_j(r) = h \times n_j(r)\, cm^{-2} \tag{4}$$

where $h$ is the depth into the cloud along our line of sight. From §3.2 $h$ is found to be 0.115 pc. The number density from the model is $n_j$ for the $j^{th}$ molecule.

[4] www.udfa.net

In most cases observations of diatomic molecules are presented as column densities. In cases where the observed surface brightness is the quantity reported, we have converted it into a column density using the assumption that the line is optically thin and the energy levels of the molecule are in LTE. Under these assumptions the relation between column density $N_T$ and surface brightness $\langle I \rangle$ averaged over frequency $v$ is given by Miao et al. (1995)

$$N_T = \frac{2.04\langle I\rangle}{\theta_a \theta_b}\left[\frac{B(T_{rot})}{B(T_{rot}) - I_{back}}\right]\frac{S_f Q_{rot}\exp\left(\frac{E_u}{T_{rot}}\right)}{g_l g_k S \mu^2 \upsilon^3} \times 10^{20}\, cm^{-2} \tag{5}$$

where $Q_{rot}$=$2kT_{rot}/hv$, and $B(T_{rot})$ is the Plank function at a temperature $T_{rot}$ with $I_{back}$ equal to the background continuum. In the analysis done by Young Owl et al. (2000) the conversion from $I$ to $N$ assumed that $T_{rot}$ = $T_{kin}$ = 100K, a value similar to the CR enhanced model. This is approximately the observed temperature of the $^{12}$CO gas. When it is not possible to derive $T_{rot}$, $T_{kin}$ is often used. This underestimates the true column density if the line is subthermally populated as occurs when the gas density is below the critical density of the line.

Figure 7a shows the $CO^+$ comparison. Storzer et al. (1995) and Fuente et al. (2003) both measured $N(CO^+)$ as a function of depth into the Bar along the same line of sight as the other molecules measured by Tielens et al. (1993) and Tauber et al. (1994). The $CO^+$ observations are characterized by a rise in column density with distance from the IF reaching a peak column density of $3\times10^{12}$ cm$^{-2}$ at $r$ =17±7". After the peak a gradual decrease is observed out to a distance of 40" where $N(CO^+)$ = $4.8\times10^{11}$ cm$^{-2}$. The column densities found by Storzer et al. are uncertain and may actually be higher based on uncertainties in the excitation mechanism assumed in the analysis, as explained in their paper. The $CO^+$ column density computed using the magnetic model with the canonical cosmic ray ionization rate peaks too early and is a factor 100 too low. With increasing cosmic ray ionization rate the peak column density increases and moves farther from the IF. For the enhanced cosmic ray model the peak column density is $4.85\times10^{12}$ cm$^{-2}$ and occurs at $r$ = 12". The subsequent decrease in the modeled $N(CO^+)$ follows the observations very closely. $CO^+$ is shown to be very sensitive to the presence of cosmic rays. Decreasing the ionization rate by a factor of 10 decreases the column density by a factor of 100 at $r \geq 20$". We conclude that the enhanced cosmic ray model is in good agreement with the observed $CO^+$ column densities.

The CN column density (Fig. 7b) of our enhanced cosmic ray model is equal to the observed value of $1\times10^{14}$ cm$^{-2}$ at 20" (Simon et al. 1997), and is 5 times higher than the observed value of $2\times10^{14}$ at $r$ = 30". These are decreased to 0.5 and 3.75, respectively, if we adopt a gas phase C abundance of 30 percent of the solar value (Jansen et al. 1995). In contrast our magnetic model underpredicts CN by more than a factor of 10 at 20" and by a factor of 3 at r = 30". We consider our enhanced model to match the CN observations to an acceptable level, given that the uncertainties in the molecular data result in uncertainties of up to an order of magnitude in the absolute chemical abundance of species like CN (Simon et al. 1997).

The $SO^+$ column density (Fig. 7c) is observed to steadily increase with distance from the IF reaching $8.4\times10^{12}$ cm$^{-2}$ (Fuente et al. 2003) at $r$ = 28". The computed column densities for the cases including the magnetic field without and with enhanced cosmic rays rise steeply at around $r$ = 20” to $3.4\times10^9$ $cm^{-2}$ and $1.1x10^{13}$ $cm^{-2}$ respectively, followed by a nearly constant plateau. Our cosmic ray enhanced model reproduces the observed peak column density, although additional $SO^+$ emission is observed near the IF and may be due to the background molecular gas.

The peak measured column densities of SO (Fig. 7d) are $6\times10^{14}$ $cm^{-2}$ at 28” (Jansen et al 1995). Taking beam dilution into account, this corresponds to a factor of 6 greater than the predicted value, which is reasonable agreement.

The observed column density of CS (Fig. 7e) is $5\times10^{13}$ $cm^{-2}$ at $r$ = 20” and increases to $5\times10^{14}$ $cm^{-2}$ at 30” (Simon et al. 1997). Hogerheijde et al. (1995) find $N$(CS) = $1.5\times10^{15}$ $cm^{-2}$ at 28” for a similar position. The profiles of our magnetic model and enhanced cosmic ray model have nearly the same value as the observations at $r$ = 20”, however these profiles are not convolved with the resolution of the observations which is very important in this case due to the rapid rise in the enhanced cosmic ray model at that point. If the resolution were taken into account, the enhanced cosmic ray model would have a column density ~$2\times10^{15}$ $cm^{-2}$ or 40 times the observed value, while the magnetic model would match the observed column density to within a factor of 2. At 30” the situation is the same with the enhanced cosmic ray model overpredicting CS by a factor of 40. These conclusions are tenuous given the large range in assumed S abundance relative to H present in the literature. For example Jansen et al. (1995) and Simon et al. (1997) found the gas phase S abundance relative to H to be $2\times10^{-7}$, while Young Owl et al. (2000) assumed $7.9\times10^{-6}$. If S/H in the PDR is changed to $2\times10^{-7}$, the abundance assumed by Simon et al. (1997), our modeled N(CS) would drop to $1.25\times10^{15}$ at $r$ = 28”, which is 0.83 times the observed value. Thus the CS column density taken by itself could easily be adjusted to fit the observations. However if the lower abundance of S that reproduces CS is used, the predicted SO and $SO^+$ become many orders of magnitude too faint. Therefore, the predicted ratio of CS to SO and $SO^+$ does not match the observations.

SiO observations (Fig. 7f) are available for three areas across the Bar (Schilke et al. 2001). The observations along their cut labeled “Bar-CO” lie in the same region of the Bar that we are studying. However, the exact location of the observations is unimportant because the measurements are statistically consistent with a constant value near $2\times10^{12}$ $cm^{-2}$ (Schilke et al. 2001). All of our models are in disagreement with this result. Our enhanced cosmic ray model predicts a column density equal to $1\times10^{13}$ $cm^{-2}$, a factor of 5 larger than observed at depths between 10” & 20”. Then $N$(SiO) sharply increases to $4\times10^{14}$ $cm^{-2}$. Any decrease in the cosmic ray density by a factor greater than 10 results in an underprediction of SiO for depth shallower than 20”, but even with the canonical Galactic value used in the magnetic model, there is an overprediction by at least an order of magnitude at depth greater than 25”. There are two explanations that may account for the discrepancy. First the gas phase Si abundance is likely to be depleted by at least an order of magnitude in the PDR compared with the ionized $H^+$ region (Schilke et al 2001). However the rate of depletion would have to be matched in such a way as to maintain a constant gas phase Si density with depth. The second possibility is that the emission is from an outflow in the foreground, although the velocity profiles suggest this is not the case (Schilke et al. 2001)

We conclude that for 4 of the 6 diatomic molecules shown, the addition of cosmic rays brings our model into general agreement (within a factor of 6) with the observations. For CS, our “magnetic model” agrees well with the observations and the further inclusion of enhanced cosmic ray heating hurts the agreement, but this depends on the fraction of S depleted onto grains in the molecular region. For SiO, none of our models agree with the observations. We stress that we did not use the molecular column densities discussed in this subsection as constraints when we fit our models to the observations. Rather, we are using them as *ex post facto* tests of how well our simulations of the inter-clump medium reproduce a wider body of data. We should also again mention that we did not do a formal calculation of the predicted emission from these molecules. Some transitions, of $CO^+$ for example, may be significantly sub-thermally populated and would therefore produce little emission and would have to be attributed to unmodeled clumps. Still Simon et al. (1997) found that the CN and CS emission could be characterized by a diffuse gas with a density of 1-4x$10^5$ cm$^{-3}$ despite the critical densities of CN N=3→2 and CS J=7→6 being 9x$10^6$cm$^{-3}$ and 3x$10^7$ cm$^{-3}$, respectively. This is comparable to our peak density of 9x$10^4$ cm$^{-3}$.

While some moderately complex molecules are included in the UMIST database and therefore in our computations, the predictions for anything more complex than diatomic molecules are highly uncertain. There are two sources of uncertainty. The first is in the rate coefficients for the many processes that lead to the formation or destruction of a species. The systematic errors in the rates are not possible to quantify. The second systematic uncertainty is in the assumptions that go into creating an equilibrium model. The comparisons presented in Röllig et al. (2007) between models computed with different codes show nearly an order of magnitude scatter in the computed number densities of these complex molecules even in cases where all models used the same fixed gas kinetic temperature and an agreed upon subset of the UMIST data set. The scatter between actual thermal equilibrium models was far worse. Because of the resulting large uncertainty in the computed column densities of these complex molecules, we will not consider them further.

**4.3 Sensitivity of final model to input parameters**

*Q(H).* During the course of this investigation, we explored the parameter space $10^{49.00} \leq Q(\mathrm{H}) \leq 10^{49.23}$ s$^{-1}$, and distance to the nebula $437 \leq d \leq 500$ pc. We could always find a placement and tilt of the Bar that would provide a good fit to the observations. It was always the case that the $H^+$ region could be described by a version of the gas pressure model, but that magnetic pressure support and cosmic ray heating were needed to also match the $H^0$ region properties. Specifically we found no combination of parameters that produced a gas pressure model in which the $H_2$ emission was displaced from the ionization front by the observed distance, nor in which the observed peak $H_2$ intensity was reproduced.

*Distance.* If the distance to the nebula were in fact 500 pc (the distance adopted by Wen & O'Dell 1995 and many other authors), $<B>$ would be affected in two ways. The projected distance of the $H_2$ emission from the IF increases from 0.021 pc to 0.024 pc. To match the larger offset a higher value of $P_{mag}/P_{gas}$ is required, with a nonlinear relation between the offset distance and $<B>$. Countering this effect, the inferred radiation pressure responsible for compressing the magnetic field also would drop, since $\theta^1$ Ori C would have to be farther away from the illuminated face. The average magnetic field in the $H^0$ region from Paper I, eq. 7 has $<B> \propto 1/R$, where $R$ is the distance to the IF. For a distance of 500 pc, $<B>$ is expected to decrease by 12 per cent. The combination of these two effects was calculated for the magnetic model without enhanced cosmic rays. The magnetic field dropped from 448μG to 435 μG.

*Inclination.* If the Bar is inclined by more than the 7 deg angle assumed here, the observed offsets would require a greater radial distance between the observed emission peaks and $\theta^1$ Ori C. Take for example a model with a calculated radial separation $\Delta x$ between two emission peaks. For a tilt angle $\phi$, the projected separation on the sky is $\Delta x' = \Delta x \cos\phi$. For increased values of $\phi$, the modeled radial separation must increase. This increased radial separation would in turn imply a lower density $H^0$ region, so that a higher magnetic field would be required to maintain hydrostatic equilibrium. The effect at 7 deg is less than 1 percent. Therefore our models with magnetic fields represent the lower limit for $<B>$ in the Bar when considering the geometry.

*Density and scattered Hα light.* The [S II] ratio found in our model is 0.05 lower than the observed value, within the 15% uncertainty in the collision rate (AGN3). Wen & O'Dell (1995) estimate that 1/3 of the surface brightness of Hα and other optical emission lines is due to light reflected from dust in the molecular cloud, which we have included in our current computed results. If this is wrong and scattered light is not such a large effect, we could compensate by making the slab somewhat larger along the line of sight.

*Turbulence.* We have considered turbulence in two ways. For all of the models presented here the turbulence was fixed to be 2 km $s^{-1}$ FWHM, consistent with observations of $H_2$ and $^{12}CO$ lines. This turbulence was counted as part of the total pressure. Another approach possible in any model with a magnetic field is to assume that $P_{mag} = P_{turb}$, motivated by equipartition arguments. In that case the turbulent velocity would vary as a function of depth and would be about 3 km $s^{-1}$ in the molecular gas. The associated increase in $P_{turb}$ then would lead to a lower magnetic field. In the enhanced cosmic ray model, the final derived cosmic ray density is in equipartition with the magnetic field, so the density of cosmic rays would also decrease. We rule out this type of model for two reasons: (1) the $^{12}CO$ emission would peak 10 arcsec farther from the ionization front than is observed; and (2) the molecular gas temperature is predicted to be only 70 K , at least 20 K colder than the observed temperature of the Bar. The offset is geometry dependent, while the temperature is not.

### 4.4 Magnetostatic equilibrium

Each of the three models assumes hydrostatic equilibrium. The left-hand column of Fig. 4 shows how the various pressure components from Eq. A2 adjust themselves to maintain this condition. As was discussed in detail in Paper I, the integrated radiation pressure from absorbed starlight steadily builds up with depth until all of the incoming photons have been used up at the bottom of the $H^0$ region. The sum of the other pressure terms must steadily rise to balance this.

It is clear from Fig. 4 that in each model there is a large residual gas pressure at the illuminated face of the cloud (depth = 0). This has to happen in these models because they describe the cloud as suddenly beginning with some gas density and temperature. The $H^+$, $H^0$ and $H_2$ regions are actually a flow from cold molecular gas through the atomic region into hot and ionized gas with the ionizing radiation gradually eating into the molecular cloud. For the case of M17 (Paper I) we found that the residual pressure at the cloud face actually is in equilibrium with the hot bubble of X-ray emitting gas that is observed to surround the ionizing stars. A similar bubble of diffuse X-ray emitting gas has recently been found to the SW of the Bar region, although the close proximity to $\theta^1$ Ori C and foreground absorption by the Veil blocks a direct view of the Bar in diffuse X-ray emission (Güdel et al. 2008). The estimated pressure from the X-Ray emitting gas was found to be roughly equal to the gas pressure of the $H^+$ region. The observed champagne flow and pressure equilibrium suggest the bubble is a leaky cavity (Güdel et al. 2008).

Our final enhanced cosmic ray model should be a realistic simulation of the Bar for a snapshot in time. The key feature is that at the same time that the radiation field from $\theta^1$ Ori C is dissociating and then ionizing the original molecular gas, the momentum carried by the photons has pushed the gas back into the molecular cloud, compressing both the gas and any magnetic field that is frozen into it. The natural and straight-forward result is that this compression is halted when the combination of gas, turbulent, and magnetic pressure has risen enough to offset the radiation pressure, so the system is in "magnetostatic equilibrium". The current quasi-equilibrium situation in the Orion Bar might represent an extrapolation of the situation described in recent MHD models (Krumholz et al. 2007) of the effects of magnetic fields on the early stages of expansion of an H II region.

### 4.5 Heating mechanisms

Figure 8 shows the relative contribution of each important heating mechanism for each of the three models. Cosmic ray heating is a minor effect in the gas pressure and magnetic field models, but in the enhanced cosmic ray model it completely dominates the heating beyond a depth of about 0.2 pc, which is well into the molecular ($H_2$) region. Cosmic ray heating accounts for 80 percent of the total heating in this region, while the remaining 20 percent is due to cosmic ray excitation of permitted FUV lines. We note that cosmic ray heating is generally thought to be responsible for heating molecular clouds (Lequeux 2005).

### 4. 6 Comparison to previous models of the Bar

*Geometry.* Almost all previous models of the Orion Bar PDR agree that the observed stratified $H^+$, PAH and $H_2$ emission must come from a diffuse gas with density 1–5x10$^4$ cm$^{-3}$, rather than from the superposition of many small optically thick clumps ( i.e. Tielens et al (1993); Tauber et al. (1994); Hogerheijde et al. (1995); Young Owl et al (2000)). The pressure of this gas is of the same order as the gas pressure at the IF inferred from the ratio of [S II] λ6716/λ6731. The geometry of the homogenous region can be estimated from the use of optically thin emission lines. With estimates of the emissivity per unit volume we have used [S II] λ6716+λ6731 to find the depth of the bar to be 0.115pc. Estimates using [O I] λ6300 and FIR continuum measurements from 20μm to 100μm lead to a similar conclusion. Since these are observations of a region close to the IF, they only directly trace the geometry at the narrow $H^+/H^0$ transition.

A more complete, 3-dimensional model of the IF over the entire nebula was constructed by Wen & O'Dell (1995), working backwards from the observed Hα surface brightness and projected positions. Their work indicated that the Bar is an upward corrugation of the main IF, which is the basic geometry that we have adopted here. We initially attempted to rather closely follow their result by describing the Bar as a surface steadily curving upward towards the observer, with $\theta^1$ Ori C at the center of curvature. We also experimented with a slab inclined at 20 deg to our line of sight, which is the tilt of the layer in the Wen & O'Dell models. In neither case were we able to reproduce the observed [S II] brightness profile and intensity simultaneously. It should be noted that Wen & O'Dell warned in their paper that their model was not expected to be very accurate in the region of the Bar. These problems led us to switch to the more nearly edge-on slab geometry described above. However, our model is still generally consistent with the basic Wen & O'Dell picture, especially if the Bar is connected to the background cloud in the way sketched in Fig. 3. We note that the computed $H_2$ profile shown in Figure 5 is narrower than the observed one, which suggests that the $H^0$ region in the Bar does have a more complicated geometry (curvature or corrugations) than we have assumed here, as Wen & O'Dell suggested. Note that the separation between $\theta^1$ Ori C and the ionization front immediately behind it is fixed by *Q(*H) and the Hα surface brightness to be 0.183 pc (this is different than the value found by Wen & O'Dell because we have adopted a different distance to the Orion Nebula, which changes *Q*(H)).

Hogerheijde et al. (1995) showed that the molecular gas also has a geometry that changes from face-on to edge-on (the Bar) and then back to face-on. Figure 13 of their paper illustrates the PDR geometry derived from $C^{18}O$. They assumed a constant abundance ratio $H_2/C^{18}O = 5\times10^6$ and $n(H_2) = 5\times10^5$ cm$^{-3}$ to estimate the volume density of $C^{18}O$. Using the measured column density of $C^{18}O = 1.3\times10^{16}$ cm$^{-2}$, the line of sight path length would be 0.6 pc. Applying our more detailed calculations and assuming $CO/C^{18}O \sim 500$ (Wilson & Rood 1994) our enhanced cosmic ray model predicts $N(C^{18}O) = 1.25\times10^{16}$ cm$^{-2}$, in good agreement with Hogerheijde et al. (1995). However the path length came out to be 0.115 pc, which is smaller than the one found by Hogerheijde et al. but is in good agreement with the path length derived above from the [S II] lines. In contrast our magnetic model has a predicted $N(C^{18}O) = 2.2\times10^{16}$ cm$^{-2}$ which would require the geometry to decrease in size by almost a factor of 2 between regions in the PDR. The similarities of the geometry between the $H^+$ region, IF and PDR therefore support our enhanced cosmic ray model.

*Interclump Region.* Our final model largely reproduces the interclump densities used by Tielens et al (1993), Tauber et al. (1994) and Young Owl et al (2000) to model the $H^0$ region. This is a factor of 5 greater than the interclump medium density proposed by van der Werf et al. (1996). The main difference in our work is in the starting point of the calculation. Previously, the observed offsets between the emission line peaks were used to establish the $H^0$ region density. Having assumed a $H^0$ region temperature of 1000 K and a constant density, Tielens et al. (1993), Tauber et al. (1994) and Young Owl et al (2000) argued that the $H^0$ region gas density was consistent with the ionized gas density for a system with gas pressure equilibrium across the IF. It is this assumption that our calculations treat in detail by including not only gas and turbulent pressure but also a pressure gradient caused by absorbed starlight. Starting from the assumption of hydrostatic equilibrium we constrained our models to match the observed properties of the $H^+$ region of the Bar. Our model using only gas pressure did not correctly describe the density and geometry of the $H^0$ region. We then added a magnetic field whose pressure is also in hydrostatic equilibrium, and the $H^0$ region density deduced by Tielens et al. then came out as a natural result.

Our models are most similar to those presented by A05 who considered a constant pressure model with an equation of state that determines the density as a function of depth, as we have, as opposed to using a constant density. Unlike van der Werf et al. (1996) for example, A05 used a single-component medium with a filling factor of unity for the $H_2$ emitting region. This was motivated by a lack of unambiguous evidence that clumping is important in the $H_2$ emitting region. A pressure of $P/k = 8 \times 10^7$ cm$^{-3}$ K was derived from the electron density and temperature at the IF. This is the same total pressure found in our models significantly beyond the IF. However, the spot measured by A05 has a higher surface brightness than the interclump region we have modeled here. Paper III shows that their $H_2$ measurements are matched by the enhanced cosmic ray model as described here but with a factor two increase in peak $H_2$ density.

The detailed study by A05 determined that a source of extra heating was required to reproduce the observed level populations of $H_2$ as well as the 100-120 K temperatures where other molecules form. They were unable to determine a realistic heating mechanism to account for this, concluding that "future modeling must address the high temperatures in the CO/$HCO^+$/$NH_3$ zone of the Orion Bar." In this paper we propose enhanced cosmic ray heating as a major source of that extra heating and present the effects of such a high cosmic ray ionization rate on the chemistry of the Bar.

*Clumping.* Tielens et al. (1993), Tauber et al. (1994) , Young-Owl et. al, (2000), van der Werf et al. (1996) and many others have also considered the effects of small (< 1 arcsec) and large scale (5–10 arcsec) regions of higher density (clumps). Clumping is directly observed in maps made in molecular lines from optically thing high levels, such as $^{12}$CO J(14-13) as well as H$^{13}$CN (Lis & Schilke 2003). We have not attempted to address this sub-structure in our model.

The Tielens et al. (1993) model for the inter-clump gas component which dominated most of the observed emission lines under-predicted the strengths of the CO (7–6) and (14–13) lines by a large factor, so these lines were attributed to emission from clumps. Our final model does in fact reproduce the observed $^{12}$CO (7–6) line strength (as well as $^{12}$CO (1–0)) to within a factor of two (Table 1). However, we under-predict the J = 14-13 line by a factor of 4, and clumps clearly are visible on direct images taken in this line (e.g. Young Owl et al. 2000), so they likely are significantly affecting the strengths of the higher-level molecular lines.  We conclude in general agreement with all the previously mentioned studies except van der Werf et al. (1996) that the UV penetration responsible for the observed stratification in the Orion Bar is dominated by a roughly homogeneous inter-clump medium, with filling factor close to unity. In our view the density concentrations comprising the clumps are just details on top of this.

Van der Werf et al. (1996) arrived at a model that had a rather lower density interclump medium ($n(H_2) = 1\times10^4$ $cm^{-3}$) in combination with clumps characterized by two different densities. A high density component with $n(H_2) \sim 10^6$ $cm^{-3}$ was deduced from photochemical models in which high densities are necessary to effectively produce hot $HCO^+$ and $CO^+$. The $^{13}CO$ J = 3-2 brightness temperatures of 40 to 50K are typically only reached in PDRs with densities of at least $10^6$ $cm^{-3}$. However, the observed CS line ratios indicate a lower-density clump component with $n(H_2) = 2.5\times10^5$ $cm^{-3}$. Our model overpredicts the CS column density (Fig. 7e) that is derived from the observed surface brightness assuming LTE and a constant gas phase S abundance from the $H^+$ region through the PDR. However, if the space density of $H_2$ is lower than the critical density of a few times $10^5$ $cm^{-3}$, the levels will be subthermally populated and the column densities deduced from the observed CS surface brightness assuming LTE will be too small. Our derived $n(H_2) = 4.6\times10^4$ $cm^{-3}$ is significantly below the critical density of CS so we conclude that the reported column densities are a lower limit. Likewise if the gas phase S abundance were lowered in the PDR our predicted CS column density would be lowered.

Neglecting clumps in our treatment is unlikely to affect our conclusions regarding the presence of a magnetic field for two reasons. First, every paper about clumps (except van der Werf et al. 1996) find the clumps to have a low filling factor (*i.e.* 0.3% clump filling factor according to Hogerheijde et al. 1995) so that they do not strongly affect the transport of the UV light responsible for the overall structure of the $H_2$ emission. Second, the studies that include clumps such as that of Young Owl et al. (2000) do not require a clumpy ridge to match the HCN and $HCO^+$ until depths greater than 20" are reached, significantly beyond the peak $H_2$ emission. Even a clumped medium does not offer a perfect explanation of the emission at this depth. As Young Owl et al. (2000) noted, significant differences in the HCN and $HCO^+$ surface brightness distributions indicates that localized variations in the production mechanisms for HCN and destruction mechanisms for $HCO^+$ are required. These may be explained by density enhancements caused by local variations in the magnetic field.

### 4.7 Is enhanced cosmic ray heating a realistic prospect?

Our best-fitting model is for the case where the cosmic ray energy density is in equipartition with the magnetic field's energy density. Equipartition occurs in the local ISM (Webber 1998) where B ~ 8μG and the energy density of cosmic rays is ~1.8eV $cm^{-3}$, although we do not know of any first-principle physical reason why this must occur. Still, the idea of equipartition has been used to argue for the existence of a high cosmic ray energy density in the Arches cluster near the galactic center (Yusef-Zadeh et al. 2007), with an energy density of $6\times10^4$ eV $cm^{-3}$. This is an order of magnitude larger than what we are suggesting here. In the Sagittarius B region the cosmic rays are thought to be enhanced by a factor of 10 over the Galactic background (van der Tak et al. 2006).

Gamma ray observations provide a limit to the cosmic ray ionization rate (Ramaty 1996). A preliminary analysis of COMPTEL satellite observations (Bloemen et al. 1994) suggested a very high gamma ray flux. Although Bloemen et al. (1999) revised this to a 2σ upper limit three times lower than the originally-claimed detection, Giammanco & Beckman (2005) used the preliminary value to find a cosmic ray ionization rate of $2–7\times10^{-13}$ $s^{-1}$, depending on the cloud mass. We rescaled their result to the revised measurement to find a 3σ upper limit on the cosmic ray ionization rate of $3\times10^{-13}$ $s^{-1}$ for $H_2$, as compared to our predicted value of $2\times10^{-13}$ $s^{-1}$. Thus the predicted gamma-ray flux from our enhanced cosmic ray model is consistent with the upper limit detected by COMPTEL.

These high energy particles in the presence of magnetic fields also produce synchrotron radiation, so we next examine whether they are ruled out by radio continuum observations. The total power emitted per unit volume for a power law distribution of particles is, from chapter 6 of Rybicki and Lightman (2004),

$$\varepsilon_{tot}(\upsilon) = \frac{\sqrt{3} q^3 C B \sin\alpha}{mc^2(p+1)} \times \Gamma\left(\frac{p}{4}+\frac{19}{12}\right) \times \Gamma\left(\frac{p}{4}-\frac{1}{12}\right) \times \left(\frac{2\pi m c \upsilon}{3qB\sin\alpha}\right)^{-\frac{(p-1)}{2}} \qquad (6)$$

where $B$ is the strength of the magnetic field, m the electron mass, $c$ the speed of light, and $\alpha$ is the angle between the electron velocities and the magnetic field. $C$ and $p$ parameterize the distribution of the volume density of relativistic electrons $n_{CR}$ as a function of energy E

$$n_{CR} = C \times E^{-p} \qquad (7)$$

where $p \sim 2.4$ as measured from the spectra of radio-loud active galaxies (Kellerman 1966) and $C$ is a normalization constant. We have assumed an initially tangled magnetic field. If the field were still tangled the synchrotron emission would be isotropic. However the field would become less tangled as it is swept up with the gas. This will result in beaming that could increase or decrease the observed flux depending on the exact orientation of the magnetic field relative to our line of sight. If to first order, we assume the radiation will still be fairly isotropic, the predicted surface brightness due to 20 cm synchrotron emission from the PDR for our enhanced cosmic ray model is

$$S_{sync} = \frac{\varepsilon}{4\pi} \times L = 1.08 \times 10^{-27}\, erg\ s^{-1} cm^{-2} arc\sec^{-2}\ Hz^{-1}. \qquad (8)$$

Continuum measurements have been made of the Orion Nebula using both single dish and interferometric radio telescopes at 20 and 2cm (Felli et al 1993). The single dish measurements provide a total flux but do not have adequate angular resolution to resolve the Bar. Interferometric observations using the VLA C and D configurations with a beam size of 28” show that the observed surface brightness at the same position in the Bar is $1.35\times10^{-26}$ erg $s^{-1}$ $cm^{-2}$ $arcsec^{-2}$ $Hz^{-1}$ at 20cm. This is a lower limit to the true flux because the total interferometric observation contains only 60 percent of the single dish flux. Assuming that this lower limit is the true value, the predicted synchrotron radiation is 8% of the observed value and therefore is consistent with the observations. The observed flux is dominated by thermal emission.

A further question is how long an enhanced rate of cosmic ray heating could go on for, even if the cosmic ray density does become large at some point in time. With no inflow of new cosmic rays, the rate at which cosmic rays heat the gas is the same as the rate at which the overall cosmic ray energy density is decreasing, and this can be used to determine a time scale. An analytical form of the cosmic ray heating rate can be found in Tielens & Hollenbach (1985). Assuming that the cosmic rays are trapped in the cloud by a tangled magnetic field, we can then define the lifetime of a high-energy cosmic ray by

$$\tau_{CR} = U_{CR} / \Gamma_{CR} \qquad (9)$$

where $U_{CR}$ is the energy density of cosmic rays and $\Gamma_{CR}$ is the heating rate from Tielens and Hollenbach. This lifetime depends on the $H_2$ and $H^0$ density, but is independent of the cosmic ray density. For the gas density $n_H = 4.6\times10^5$ $cm^{-3}$ which exists in the appropriate zone of our best model, the cosmic ray lifetime is only 2000 years. This short lifetime applies to all PDRs with similar densities. The transport of cosmic rays through a magnetized partially ionized medium is a rich and complex problem with no easy solutions (see, for example, Lazarian & Beresnyak 2006; Snodin et al. 2006). The details of their motion depend on the field geometry, which is unknown in this region of Orion. However, the timescale problem we have pointed out is shared by most models of the neutral and molecular regions. The short timescale is a consequence of the hydrogen density and not the cosmic ray density.

We have shown that an enhanced cosmic ray density can account for the observed properties of the Bar. Enhanced cosmic rays will result from compression of magnetic field lines. If the field is well ordered and connected to the diffuse ISM the cosmic rays will leak out at nearly the speed of light and the CR density quickly will go back to the background value. If field lines are tangled, as assumed in our work, the CRs will be trapped and lose their energy through collisional processes. This is clearly a crucial issue. Recent studies (Indriolo et al. 2007) have shown that the cosmic ray ionization rate varies widely along various sight lines through the diffuse ISM with a maximum value over 1 dex larger than the accepted background value. Clearly the picture of a single Galactic cosmic ray background with ordered fields sustaining this rate is an oversimplification.

**4.8 Comparison to the magnetic field in other PDRs**

The other star forming region where this type of analysis has been performed is M17 (Paper I). The edge-on ionization front in M17 has a considerably lower gas density than does the Orion Bar (the [S II] ratio indicates $n_e$ ~560 $cm^{-3}$ in M17, as compared to $n_e$ ~3200 $cm^{-3}$ in the Orion Bar). At the same time, the magnetic fields are measured to be roughly equal in the two $H^0$ regions. This means that the ratio $P_{mag}/P_{gas}$ is much larger in M17, so that the magnetic field has a much larger (and hence more noticeable) effect. Paper I showed that the magnetic field produces a very large increase in the extent of the $H^0$ region in M17, as is observed. In the case of the Orion Bar, $P_{mag}/P_{gas} \sim 1$ in the $H^0$ region (Fig. 4) so the $H^0$ region size is increased by only about a factor of two. In addition, while the magnetic pressure also affects the region where the [S II] doublet is formed in M17, this does not really happen in Orion. None-the-less, the magnetic pressure does change the overall structure of the Orion Bar by a significant amount, and so should be included in models of it.

At a depth greater than 0.15pc into the cloud our best model predicts most hydrogen is molecular and has a density $n(H_2)=10^{4.66}$ $cm^{-3}$, and a magnetic field of 543μG. Crutcher (1999) has combined most of the available direct measurements of $|B_{los}|$ for a number of systems, including molecular cores and star forming regions such as M17, along with estimates of $n(H_2)$. Figure 9 has been adapted from Figure 1 of Crutcher (1999) to include our point for the Orion Bar. The filled circles are measured points for other systems, while the triangles represent systems for which only an upper limit on $B_{los}$ is available. The Orion point is consistent with the general trend, showing that the magnetic field strength that we have deduced for the Bar is in line with the values found in the cases where direct measurements are possible.

**4.9 The effect of radiation pressure on gas density**

In the simplest case, the Bar is an edge-on ionization front where the observed increase in surface brightness can be attributed to the depth along the line of sight. However, this cannot be the whole story. The maps of the measured [S II] λ6716/λ6731 ratio shown by both Pogge et al. (1992) and García-Díaz & Henney (2007) show that the Bar is also a density enhancement over the regions to either side of it. We suggest that this increase in density is due to the Bar being more face-on to the light from $\theta^1$ Ori C than is the surrounding ionized cloud face, so that radiation pressure due to starlight is greater in the Bar. To illustrate this, we perturbed the momentum (as set by $\Phi(H) \propto Q(H)$) at the illuminated face upwards and downwards by a factor of 2. Figure 10 shows that the density at the ionization front responds roughly according to $n_H \propto \Phi(H)$, indicating that in the $H^+$ zone the momentum $p \propto \Phi(H)$ is balanced by gas pressure.

This effect, the changing flux of momentum in the ionizing radiation field, may be what drives the density variations across many H II regions. It appears to occur over the Orion Nebula as a whole. For example, in a strip reaching radially to the W of $\theta^1$ Ori C that is covered by the long-slit spectrum of BFM91, the electron density falls off steadily as a function of the distance $r$ from $\theta^1$ Ori C approximately as $n_e \propto r^{-1.7}$ (see their Fig. 7), consistent with the dilution of the incident momentum entering the cloud per unit area along this strip. Wen & O'Dell (1995) find $n_e \propto r^{-1.63}$ from a more comprehensive plot of $n_e$ vs. $r$ covering the two-dimensional face of the Orion Nebula (their Fig. 6). While this is not quite as rapid a falloff as the expected $r^{-2}\cos(\theta)$ dependence (where $\theta$ is the angle at which the ionizing radiation strikes the IF), it certainly suggests that the decrease in momentum flux is an important factor. Radially decreasing density gradients also are found in a number of additional H II regions, both Galactic (Copetti et al. 2000) and extragalactic (Castañeda et al. 1992). These may be other cases in which the ionization front forms a background sheet behind the ionizing star(s), similar to the situation in Orion.

**5. Conclusions**

We simultaneously modeled the Orion Bar's $H^+$ and PDR regions in order to find out which physical processes are important in determining the overall structure. We assumed that the Bar is in hydrostatic equilibrium throughout; while not likely to be strictly true, this should be a fair approximation. We compared models that include a magnetic field that is well-coupled to the gas to models without magnetic fields and developed a final best-fitting model that includes both magnetic fields and an enhanced cosmic ray density.

Our final model reproduces the observed surface brightness profiles, in the sense of both their position and the actual brightness, of the H$\alpha$, [S II], $H_2$ and $^{12}$CO emission lines and also the [S II] $\lambda$6716/$\lambda$6731 ratio, and the strengths of [O III], [N II], [O I], [Si II], [C II] and additional CO emission lines. This is achieved by varying the location of the illuminated face and the gas density at the face, the tilt of the IF and its depth along the line of sight, the magnetic field strength, and the cosmic ray density. These emission lines come from what are really three very different regions (ionized, atomic and molecular) associated with star formation. Self consistently modeling all the zones together provides us with additional constraints that are not available when treating either the $H^+$ or the combined $H^0$ and molecular zones as independent entities.

We find that in the case of the Bar, magnetic pressure plays an important role in the $H^0$ region, providing about half of the total pressure support. The evidence for this is the offset of the $H_2$ emission peak from the position of the ionization front as marked by the [S II] emission peak, which is mostly a measurement of the gas density in the $H^0$ region. Since in hydrodynamic equilibrium the magnetic pressure offsets some of the need for gas pressure, the result is a lower gas density. For this model we find $\langle B \rangle = 438\mu$G in the $H^0$ region.

The above model still did not provide a very good fit to the location and brightness of the $^{12}$CO J=1–0 emission peak, nor did it reproduce the extended $H_2$ emission reaching into the deeper regions of the cloud. We find that these features can be explained if there is an associated enhancement in the density of cosmic rays by the factor expected if the cosmic ray and magnetic field energy densities are in equipartition, with $\langle B \rangle = 516\mu$G in the $H^0$ region and the cosmic ray density $10^{3.6}$ times higher than the Galactic background value. Although we don't know exactly how this equipartition would occur, and there may be problems about the timescales over which these cosmic rays lose their energy, this is the same situation that we found in M17. This suggests that an increase in the cosmic ray density may be a natural consequence of the compression of magnetic fields frozen into these gas clouds. We find that an enhanced abundance of cosmic rays in a diffuse inter-clump medium can explain the observed $CO^+$ surface brightness profile as well as the surface brightness profiles of a number of other molecules, although some molecular lines clearly do come from clumps.

We note that the initial magnetic field (before compression) needed to create such a scenario is very close to the average field for the Galaxy, suggesting that magnetic fields may be important in many similar regions of star formation. In both this current model of the Orion Bar and in our earlier model of M17, the basic concept is that the momentum carried by radiation and winds from the newly formed stars compresses the surrounding gas until enough pressure builds up inside the gas cloud to resist (i.e. until an approximate hydrostatic equilibrium is set up). If the gas starts out with a weak magnetic field, that field can be compressed along with the gas until magnetic pressure becomes an important factor and magnetostatic equilibrium is established. The enhancement of the cosmic ray density would then just be an additional side effect. We believe that this is a very straight-forward, cause-and-effect description of some of the key processes that shape the gas clouds that are found in star-forming regions of all sizes.

EWP and JAB gratefully acknowledge support from NSF grant AST-0305833. GJF thanks NSF for support through AST-0607028, NASA for support through NNG05GD81G and STScI for support through HST-AR-10653.

## Appendix A. The gas equation of state

The equation of state is the relationship between the gas density and pressure. The terms we include in the equation of state are described here.

There are three broad classes of simulation codes that could be applied to a region such as the Orion molecular cloud. Hydrodynamics codes such as Zeus (Hayes et. al 2006) or Gadget (Springel 2005) will follow the gas dynamics in detail, often in three dimensions and sometimes with a treatment of the magnetic field. This class of codes generally does not do the atomic physics in detail but rather treats thermal and ionization processes with generalized fits to universal functions and neglect the radiative transfer. Radiative transfer codes such as ATLAS (Kurucz 2005) or Phoenix (Hauschildt et al. 1997) will do the radiative transfer with great care but at the expense of the atomic physics and dynamics. Finally, plasma simulation codes such as Cloudy, which we use in this paper, treat the atomic, molecular, and emission physics with great care but do the dynamics and radiative transfer with more approximate methods.

All three classes of codes are trying to do the same thing, a true simulation of what occurs in nature, but are limited by available computers and coding complexity. All are being improved in the areas in which they are weak but we are still many years away from being able to perform a true simulation of the spectral emission of a magnetized molecular cloud with an advancing ionization front.

In this paper we model the $H^+$ / $H^0$ / $H_2$ layers as a hydrostatic atmosphere. This is clearly a simplification – the layers are actually a flow from cold-molecular into hot-ionized regions. For a D-critical ionization front the ram pressure, the pressure term due to the motions of the gas, will equal the gas pressure once the gas has attained its full motion (Henney et al 2005b; Henney 2007). This term is not present in a static geometry, so the pressure we use may be off by as much as a term equal to the gas pressure. In most of the cloud the gas pressure is only a fraction of the total pressure, which includes terms from turbulence, radiation pressure, and the magnetic field. The uncertainty introduced by the hydrostatic approximation is likely to be of the same order as uncertainties in the chemistry network or the grain properties.

We treat magnetic pressure as a scalar which is added to the gas and radiation pressures. This is formally correct if the field is highly disordered. For an ordered magnetic field the forces acting on a charged particle produce a directed motion rather than a scalar pressure term. We know that the magnetic field in the Veil, the atomic layer of gas in front on the Orion Nebula (Abel et al. 2005), is ordered since a disordered field would produce no net Zeeman polarization. The field strength and geometry across the Bar are unknown. The approach we take is simple but reasonable given the uncertainties and complexities.

The magnetic field is computed assuming flux freezing, that is, that the field and gas are well coupled. The field at any place in the cloud is related to the field at the illuminated face of the cloud by the ratio of densities. The gas density at the ionization front is in turn constrained by the observed [S II] I(λ6716)/I(λ6731) intensity ratio. As is explained in detail in Paper I, the magnetic field is assumed to scale with the gas density according to the relation

$$B = B_0 \times \left( \frac{n}{n_0} \right)^{\gamma/2} . \qquad \text{(A1)}$$

Here $B_0$ and $n_0$ are the magnetic field and the gas density at the illuminated face of the cloud and $\gamma$ depends on the geometry of the system. For spherical collapse $\gamma$ is 4/3 while $\gamma = 2$ describes the 2D compression of a shell. We choose $\gamma = 2$ on the assumption that the Bar is an edge-on section of the general shell of material which has been swept up and compressed by the radiation pressure from $\theta^1$ Ori C. This is similar to the situation in M17, for which this choice was justified in some detail in Paper I.

The total pressure $P_{tot}$ is then given by

$$P_{tot}\left(r\right) = nkT + \frac{B^2}{8\pi} + P_{turb} + P_{lines} + P_{stars}\left(r\right) \qquad \text{(A2)}$$

This is the equation of state assumed in BFM91, who ignored the magnetic and turbulent pressure terms. They called this a constant pressure model, and Cloudy continues to do so today, although hydrostatic is a better term. The first term in Eq. A2 is thermal gas pressure, the second is the magnetic pressure, $P_{turb}$ is the pressure from non-thermal turbulent motions, $P_{lines}$ is the radiation pressure due to trapped emission lines (mainly Lyα), and $P_{stars}$ is the net pressure resulting from the absorption of starlight. This last term is given by

$$P_{stars}(r) = \int_{r_1}^{r} a_{rad} \rho \, dr \qquad \text{(A3)}$$

This is equal to zero at the illuminated face, where $r = r_1$, and then grows with depth, reaching its full value near the ionization front, where $r = R_{\rm H}$. The total outward force is approximately given by the total momentum in ionizing radiation,

$$P_{stars} = \frac{Q\left(\mathrm{H}^0\right)\left\langle h\nu \right\rangle}{4\pi R_{\rm H}^2 c} . \qquad \text{(A4)}$$

The local pressure is not actually constant, although there is no net force acting on the gas, since the $P_{stars}$ term increases with increasing depth. Paper I gives a more detailed description of these effects.

**References**

Abel, N.P., Brogan, C.L., Ferland, G.J., O'Dell, C.R., Shaw, G. & Troland, T.H. 2004, ApJ, 609, 247

Abel, N.P., Ferland, G.J., Shaw, G. & van Hoof, P.A.M. 2005, ApJS, 161, 65

Abel, N.P., Ferland, G.J., O'Dell, C.R., Shaw, G. & Troland, T.H. 2006, ApJ, 644, 344

Abel, N.P. & Ferland, G.J. 2006, ApJ, 647, 367

Allers, K. N., Jaffe, D. T., Lacy, J. H., Draine, B. T. & Richter, M. J. 2005, ApJ, 630, 368 (A05)

Bakes, E. L. O. & Tielens, A. G. G. M. 1994, ApJ, 427, 822

Baldwin, J. A., Ferland, G. J., Martin, P. G., Corbin, M. R., Cota, S. A., Peterson, B. M. & Slettebak, A. 1991, ApJ, 374, 580 (BFM91)

Balick, B., Gammon, R. H. & Hjellming, R. M. 1974, PASP, 86, 616

Black, J. H. & van Dishoeck, E. F., 1987, ApJ, 322 412

Bloemen, H. et al 1994, A&A, 281, L5

Bloemen, H. et al., 1999, ApJ, 521, L137

Brogan, C.L. & Troland, T.H. 2001, ApJ, 560, 821

Brogan, C. L., Troland, T. H., Abel, N. P., Goss, W. M. & Crutcher, R. M. 2005, ASP Conference Series, 343, 183

Brogan, C. L., Troland, T. H., Roberts, D. A., & Crutcher, R. M. 1999, ApJ, 515, 304

Castañeda, H.O, Vilchez, J.M. & Copetti, M.V.F 1992, A&A, 260, 370

Crutcher, R. M., 1999, ApJ, 520, 706

Copetti, M.V.F., Mallmann, J.A.H., Schmidt, A.A. & Castañeda, H.O. 2000, A&A, 357, 621

Draine, B. T. & Bertoldi, F. 1996, ApJ, 468, 269

Draine, B. T. 1978, ApJS, 36, 595

Draine, B. T. et al. 2007, ApJ, 663, 866

Feigelson, E. et al. 2005, ApJS, 160, 379

Felli, M., Churchwell, E., Wilson, T. L. & Taylor, G. B., 1993, A&AS, 98, 137

Ferland, G.J. 2001, PASP, 113, 41

Ferland, G. J. Korista, K.T. Verner, D.A. Ferguson, J. W. Kingdon, J.B. & Verner, E.M. 1998, PASP, 110, 761

Fuente, A., Rodrıguez-Franco, A., Garcıa-Burillo, S., Martın-Pintado, J. & Black, J. H 2003, A&A, 406, 899

García-Díaz, Ma. T. & Henney, W. J. 2007, AJ, 133, 952

Giammanco, C. & Beckman, J. E., 2005, A&A, 437, L11

Güdel, M., Briggs, K.R., Montmerle, T., Audard, M., Rebull, L. & Skinner, S.L. 2008, Science, 319, 309

Hanson, M. M., Howarth, I. D. & Conti, P.S. 1997, ApJ 489, 698

Hauschildt, P. H., Baron, E. & Allard, F. 1997, ApJ, 483, 390

Hayes, J. C., Norman, M. L., Fiedler, R. A., Borden, J. O., Li, P. S., Clark, S. E., ud-Doula, A. & Mac Low, M-M. 2006, ApJS, 165, 188

Henney, W. J. 2007, Diffuse Matter from Star Forming Regions to Active Galaxies, Hartquist, T.W., Pittard, J.M., Falle, S.A.E.G (Dordrecht : Springer), 103

Henney, W. J., Arthur, S. J. & García-Díaz, Ma. T. 2005a, ApJ, 627, 813

Henney, W. J., Arthur, S. J., Williams, R. J. R., Ferland, G. J. 2005b, ApJ, 621, 328

Hirota, T. et al. 2007, PASJ, 59, 897

Hogerheijde, M.R., Jansen, D.J. & van Dishoek, E. F. 1995, A&A, 294 792

Indriolo, N., Geballe, T.R., Oka, T. & McCall, B.J. 2007, ApJ, 671, 1736

Jansen, D. J., Spaans, M., Hogerheijde, M. R. & van Dishoeck, E. F. 1995, A&A, 303, 541

Kellerman, K.I. 1966 ApJ, 146,621

Kristensen, L.E., Ravkilde, T.L., Field, D., Lemaire, J. L. & Pineau Des Forêts, G. 2007, A&A, 469, 561

Krumholz, M.R., Stone, J.M. & Gardiner, T.A. 2007, ApJ, 671, 518
Kurucz, R. L., 2005, MSAIS, 8, 189
Kurucz, R. L., 1979, ApJS, 40, 1
Lazarian, A. & Beresnyak, A. 2006, MNRAS, 373, 1195
Lequeux, J. 2005, The interstellar medium, Translation from the French language edition of: Le Milieu Interstellaire by James Lequeux, EDP Sciences, 2003 Edited by J. Lequeux. Astronomy and Astrophysics Library, Berlin: Springer, 2005.
Leurini, S. et al. 2006, A&A, 454, L47
Miao, Y., Mehringer, D.M., Kuan, Y-J & Snyder, L.E 1995, ApJ, 445, L59
O'Dell, C.R. 2001, ARA&A, 39, 99
O'Dell, C. R & Doi, T. 1999, PASP, 111, 1316
O'Dell, C.R., Walter, D.K. & Dufour, R.J. 1992, ApJL, 399, L67.
Osterbrock, D.E. & Ferland, G.J. 2006, Astrophysics of Gaseous Nebulae and Active Galactic Nuclei, 2nd Edition, University Science Books, Sausalito (AGN3)
Osterbrock, D.E., Tran, H.D. & Veilleux, S. 1992, ApJ, 389, 196
Pellegrini, E. W. et al. 2007, ApJ, 658, 1119 (Paper I)
Pogge, R.W., Owen, J. M. & Atwood, B. 1992, ApJ, 399, 147
Ramaty, R. 1996, A&AS, 120, 373
Roshi, D.A. 2007, ApJ(Letters), 658, L41
Röllig, M., et al., 2007, A&A, 467, 187
Rybicki, G. B. & Lightman, A. P. 2004, Radiative Processes in Astrophysics, 1st edition, Wiley-Interscience,Weinheim
Rubin, R. H., Simpson, J. R., Haas, M. R., & Erickson, E. F. 1991, ApJ, 374, 564
Rubin, R.H., Dufour, R.J. & Walter, D.K. 1993, ApJ, 413, 242
Schilke, P., Pineau des Forêts, G., Walmsley, C. M. & Martín-Pintado, J. 2001, A&A, 372, 291
Schleuning, D.A., 1998, ApJ, 493, 811
Shaw, G., Ferland, G. J., Abel, N. P., Stancil, P. C., & van Hoof, P. A. M. 2005, ApJ, 624, 794
Shaw, G., Ferland, G. J., Henney, W. J., Stancil, P. C., Abel, N. P., Pellegrini, E.W., Baldwin, J.A. & van Hoof, P. A. M. 2008, ApJ, (submitted) (Paper III)
Simon, R., Stutzki, J., Sternberg, A., Winnewisser, G., 1997, A&A, 327, 12
Snodin, A.P., Brandenburg, A., Mee, A.J. & Shukurov, A. 2006, MNRAS, 373, 643
Smith, L.J., Norris, R.P.F. & Crowther, P.A. 2002, MNRAS, 337, 1309
Springel, V., 2005, MNRAS, 364, 1105
Sternberg, A., Hoffmann, T.L. & Pauldrach, A.W.A. 2003, ApJ, 599, 1333.
Störzer, H., Stutzki, J. &; Sternberg, A. 1995, A&A, 296, L9
Tauber J. A., Tielens, A. G. G. M., Meixner, M. & Goldsmith, P. 1994, ApJ, 422, 136
Tielens A. G. G.M. & Hollenbach, D. 1985, ApJ, 291, 722
Tielens, A. G. G. M., Meixner, M. M., van der Werf, P. P., Bregman, J., Tauber, J. A., Stutzki, J., & Rank, D., 1993, Science, 262, 86
Vacca, W. D., Garmany, C. D. & Shull, J. M. 1996, ApJ, 460, 914
van der Tak, F.F.S. et al. 2006, A&A, 454L, 99
van der Werf, P. P., Stutzki, J., Sternberg, A., Krabbe, A., 1996, A&A, 313, 633
Wade, G.A., Fullerton, A.W., Donati, J.-F., Landstreet, J.D., Petit, P. & Strasser, S. 2006, A&A, 451, 195
Walmsley, C. M., Natta, A., Olivia, E., & Testi, L. 2000, A&A, 364, 301
Webber, W.R. 1998, ApJ, 506, 329
Wen, Z. & O'Dell, C. R. 1995, ApJ, 438, 784
Williams, J. P., Bergin, E.A., Caselli, P., Myers, P. C. & Plume. R. 1998, ApJ, 503, 689
Wilson, T.L. & Rood, R. 1994, ARA&A, 32, 191
Young Owl, R.C., Meixner, M.M., Wolfire, M., Tielens, A. G. G. M. & Tauber, J. 2000, ApJ, 540, 886

Yusef-Zadeh, F., Wardle, M. & Roy, S., 2007 ApJL, 665, 123
Zuckerman, B. 1973, ApJ, 183, 863

**Table 1**
**Observed and predicted quantities for the Orion Bar.**

| Quantity[1] | Observed (Ref[2]) | Gas Pressure | Magnetic | Enhanced Cosmic Rays |
|---|---|---|---|---|
| [S II] ratio | 0.62 | 0.59 | 0.59 | 0.59 |
| *S*([S II]$\lambda$6716+$\lambda$6731) | 5.7e-13 (1) | 5.6e-13 | 5.5e-13 | 5.6e-13 |
| *S*(H$\alpha$) | 6.6e-12 (2) | 6.5e-12 | 6.5e-12 | 6.5e-12 |
| *S*(O III) | 6.8e-12 (2) | 8.2 e-12 | 8.2e-12 | 8.2e-12 |
| *S*(N II) | 2.2e-12 (2) | 2.8e-12 | 2.7e-12 | 2.7e-12 |
| *S*($H_2$ 2.121$\mu$m) | 3.3e-15 (3) | 5.0e-15 | 3.3e-15 | 3.1e-15 |
| *S*($^{12}$CO J(1-0)) | 9.4e-18 (4) | 1.7e-17 | 8.0e-18 | 7.5e-18 |
| *S*($^{12}$CO J(7-6)) | 4.7e-15 (4) | 3.0e-15 | 4.3e-17 | 7.2e-15 |
| *S*($^{12}$CO(14-13)) | 7.1e-15 (4) | 2.0e-17 | 2.6e-21 | 2.6e-15 |
| *S*(OI 145$\mu$m) | 4.7e-14 (4) | ≤ 2.9e-13 | ≤ 1.3e-13 | ≤ 1.2e-13 |
| *S*(OI 63$\mu$m) | 9.4e-13 (4) | ≤ 4.4e-12 | ≤ 1.8 e-12 | ≤ 2.2e-12 |
| *S*(Si II 34$\mu$m) | 2.1e-13 (4) | 8.8e-13 | 3.7e-13 | 3.9e-13 |
| *S*(CII 158 $\mu$m) | 1.2e-13 (4) | 2.4e-13 | 1.2e-13 | 1.2e-13 |
| <*B*> $\mu$G[3] | Unknown | 0$\mu$G | 438$\mu$G | 516$\mu$G |

[1] Peak surface brightness *S* is in erg $s^{-1}$ $cm^{-2}$ arcsec $^{-2}$.
[2] References for peak surface brightness: (1) new SOAR observations; (2) Wen and O'Dell 1995; (3) Young Owl et al. 2000; and (4) Tauber et al. 1994.
[3] <*B*> is weighted by $T_{spin}/\ n(H^0)$.

**Table 2**
**Comparison of face-on predicted optical lines to BFM spectrum**

| Emission Line | Observed[1] | Enhanced Cosmic Ray Model[2] | Model/Observed |
|---|---|---|---|
| [O II] 3727 | 1.246 | 1.43 | 1.14 |
| [Ne III] 3869 | 0.165 | 0.366 | 0.22 |
| Hγ | 0.460 | 0.467 | 1.02 |
| He I λ4471 | 0.044 | 0.045 | 1.04 |
| [O III] λ4959 | 1.052 | 0.955 | 0.91 |
| [O III] λ5007 | 3.144 | 2.875 | 0.91 |
| [O I] λ5577[3] | 0.003 | 0.0002 | 0.06 |
| [N II] λ5755 | 0.007 | 0.008 | 1.10 |
| He I λ5876 | 0.133 | 0.136 | 1.02 |
| [O I] λ6300[3] | 0.025 | 0.010 | 0.41 |
| [O I] λ6363[3] | 0.007 | 0.003 | 0.50 |
| Hα | 2.960 | 2.893 | 0.98 |
| [N II] λ6584 | 0.548 | 0.588 | 1.07 |
| He I λ6678 | 0.034 | 0.035 | 1.04 |
| [S II] λ6725 | 0.070 | 0.113 | 1.61 |
| [S II] λ6717 | 0.026 | 0.042 | 1.61 |
| [S II] λ6731 | 0.051 | 0.071 | 1.39 |
| [S II] λ6731/λ6717 | 1.642 | 1.663 | 1.02 |
| He I λ7065 | 0.059 | 0.085 | 1.44 |
| [Ar III] λ7751 | 0.036 | 0.049 | 1.36 |
| [S III] λ9069+λ9532 | 1.705 | 1.600 | 0.94 |
| [S III] 9532 | 1.452 | 1.140 | 0.79 |

[1]Line strengths from BFM91, relative to Hβ.
[2]Model predictions are for a face-on observation, relative to Hβ.
[3]Observed line is blended with night sky emission.

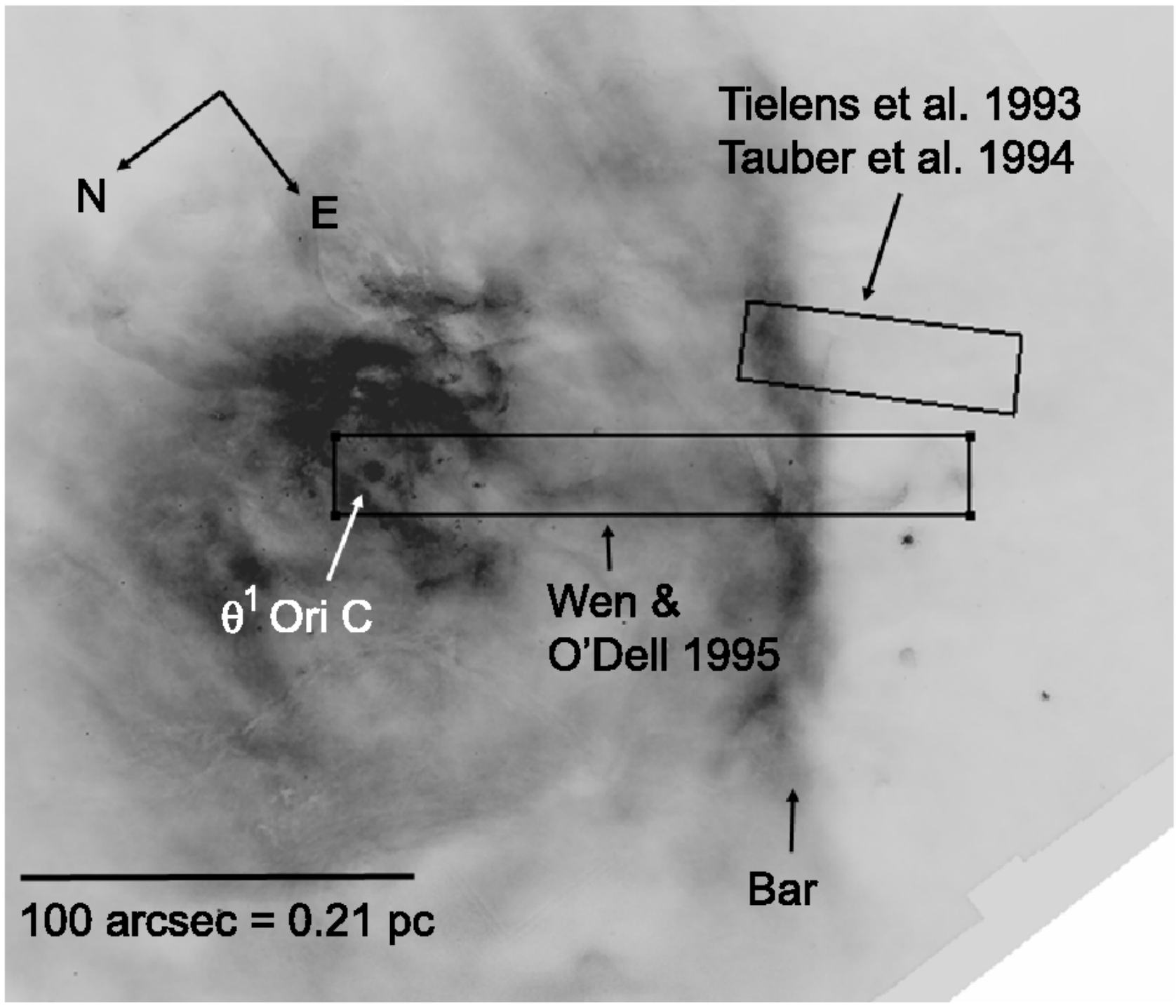

Figure 1. Positions of data across the Orion Bar used in this analysis, shown superimposed on a dereddened Hα image provided by C.R. O'Dell. The lines included for each cut are: Wen & O'Dell (1995), Hα; Tauber et al. (1994), $H_2$ and $^{12}CO$. The image is rotated so that the ionizing radiation strikes the Bar from the left, the same as in Figs. 2, 3, 4, 5, 7 and 8.

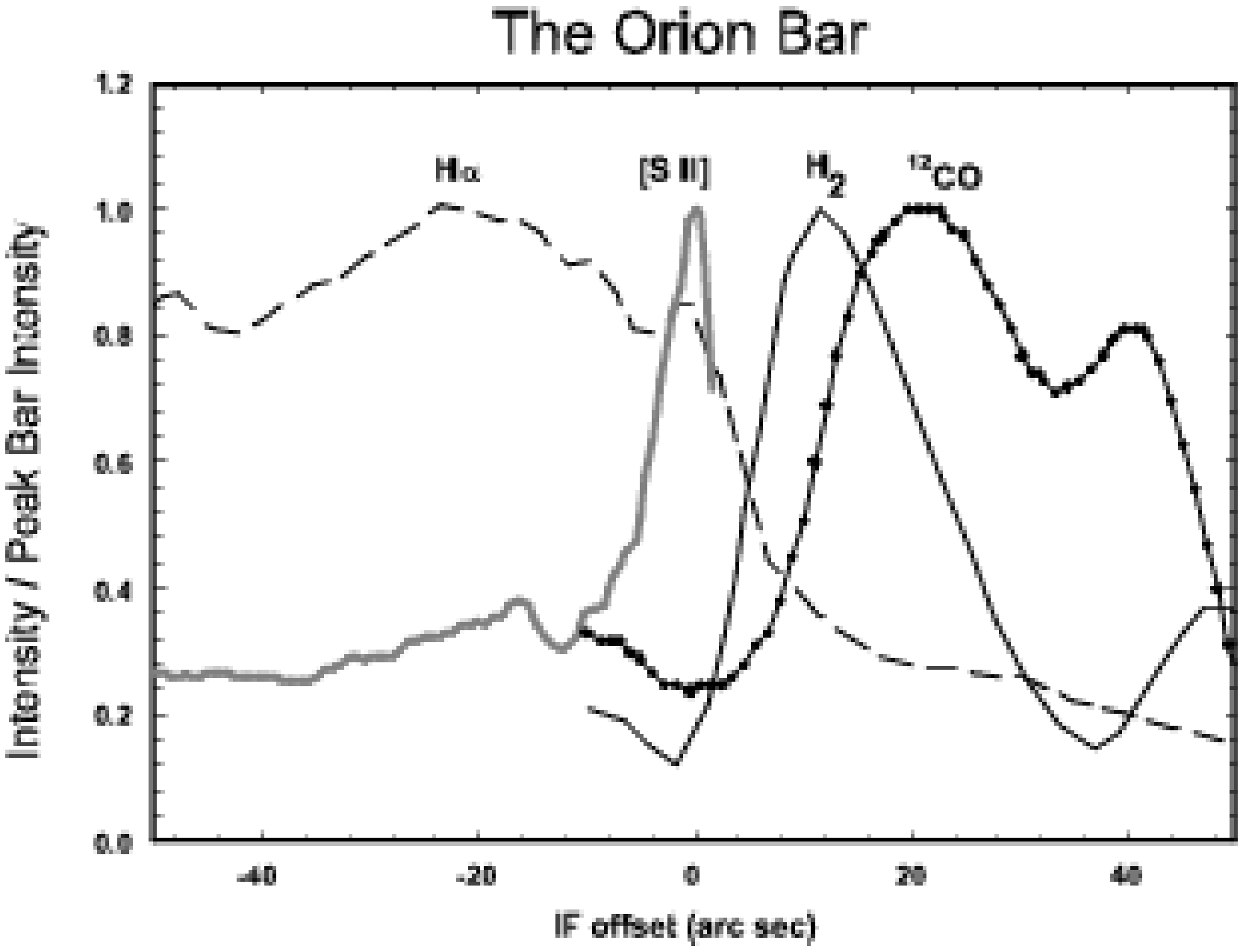

Figure 2 Observations of the Orion Bar from Tielens et al. 1993 ($^{12}CO$ ), Wen and O'Dell 1995 (Hα), Young Owl et al. 2000 ($H_2$) and this paper (SII [λ6716+λ6731]), all relative to the IF defined by the peak in the [S II] emission. $\theta^1$ Ori C is to the left at –111 arcsec. There is clear stratification indicating an ionized region viewed nearly edge-on.

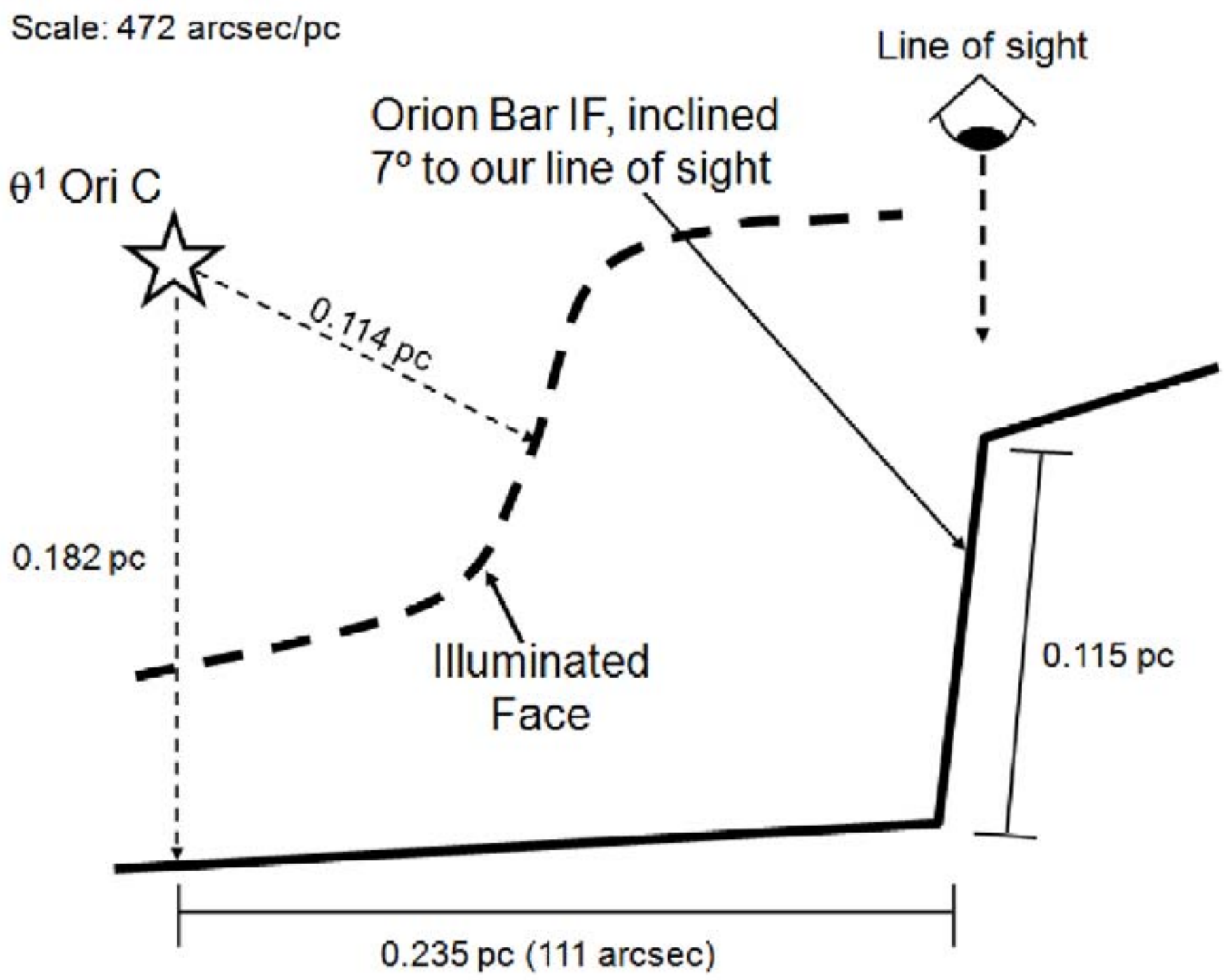

Figure 3. The geometry derived from the [S II] emission at the ionization front. The Bar is well represented by a slab 0.115 pc long inclined at 7 deg to the viewing angle.

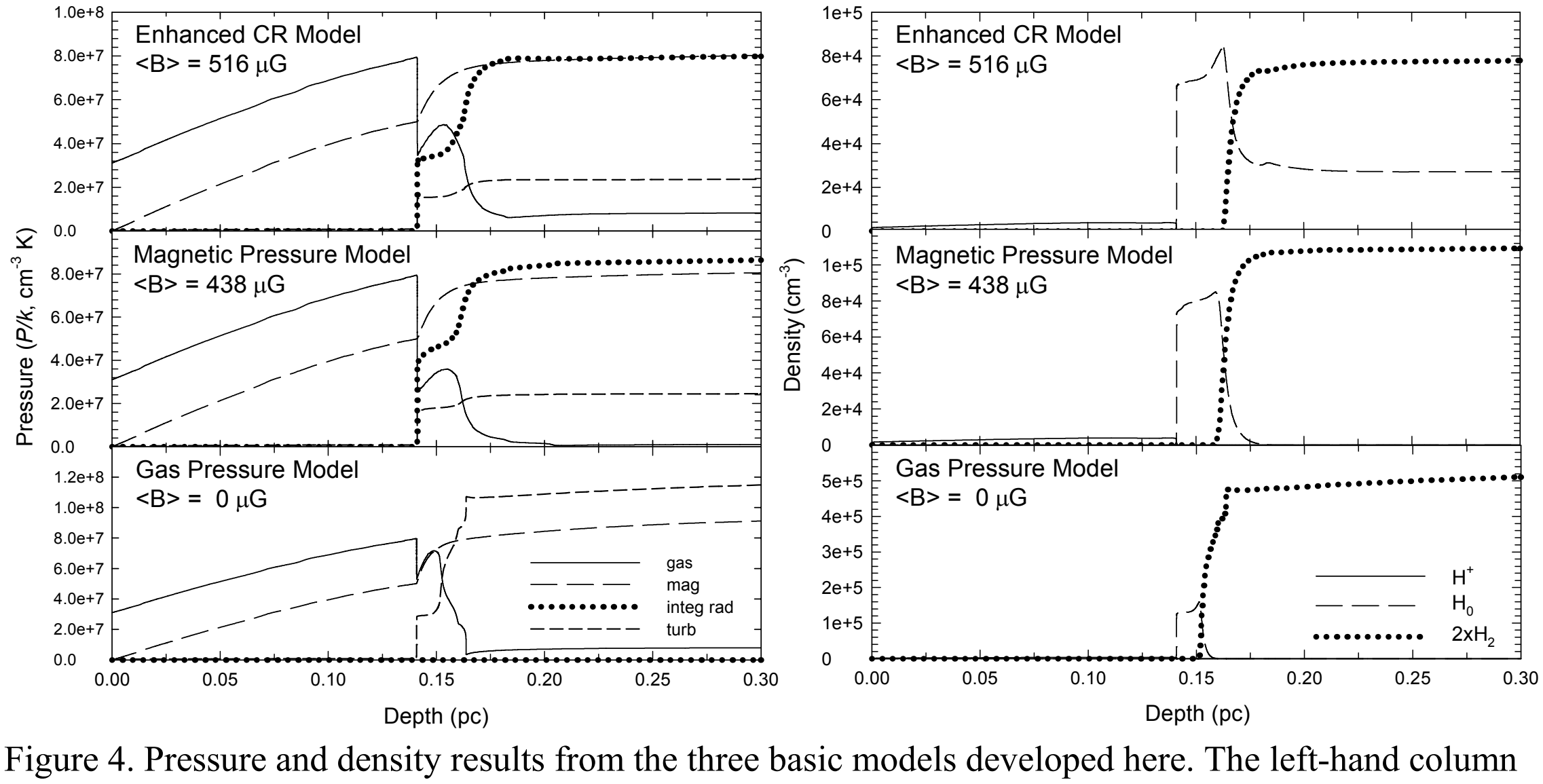

Figure 4. Pressure and density results from the three basic models developed here. The left-hand column shows the various pressure components (gas, magnetic, absorbed radiation from stars, turbulence) as a function of depth into the cloud from its illuminated front surface. The right-hand column shows the number density of H atoms in the $H^+$, $H^0$ and $H_2$ zones, as a function of depth, so that the pressures shown on the left can easily be related to specific zones in the model.

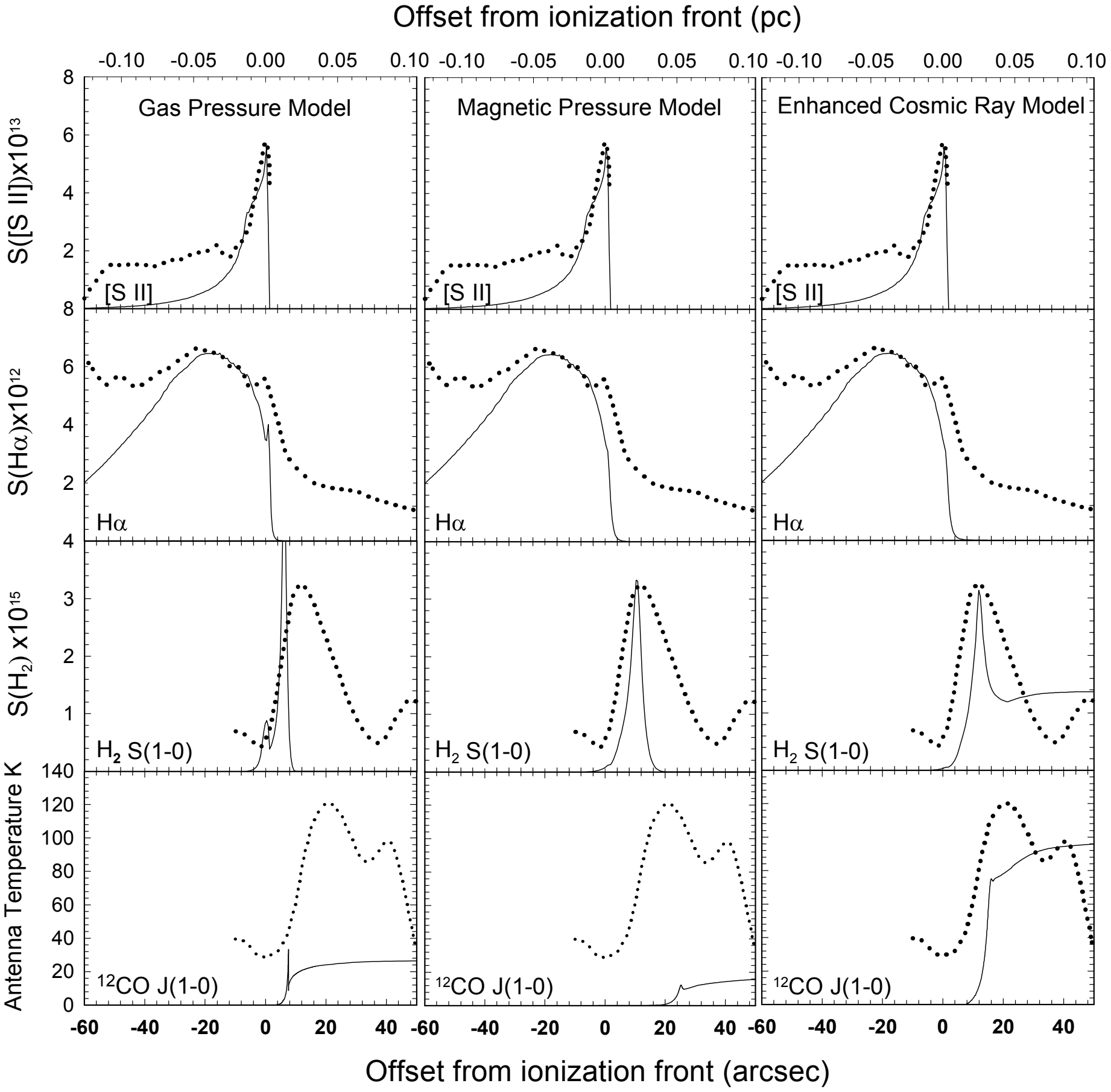

Figure 5.The surface brightness distributions in key emission lines, in units of erg s$^{-1}$ cm$^{-2}$ arcsec $^{-2}$ or antenna temperature, as computed for the three basic models (solid lines), compared to the observed distributions (dotted lines).

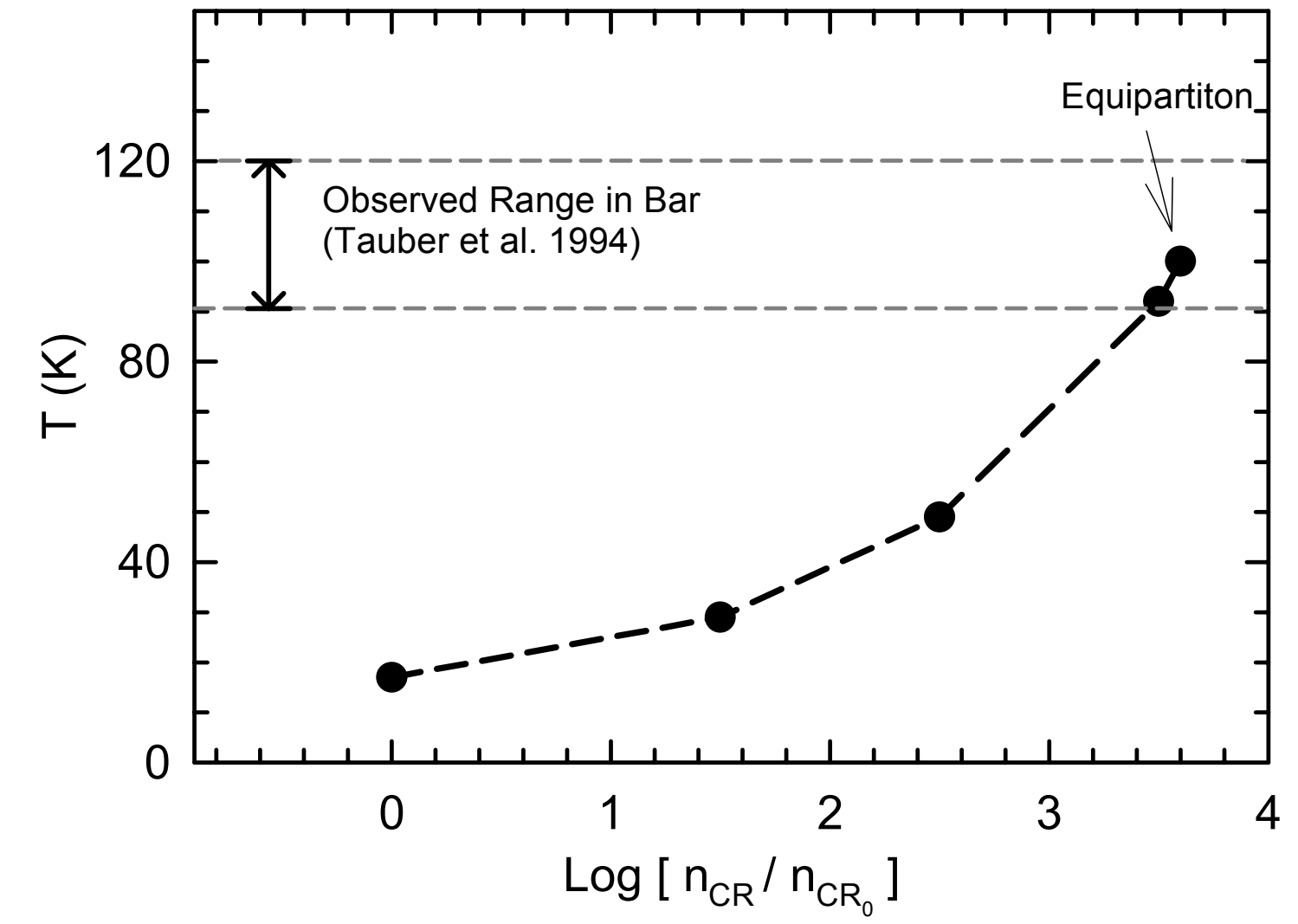

Figure 6. Predicted $^{12}$CO brightness temperature as a function of cosmic ray density normalized by the Galactic background cosmic ray density $n_{CRo}$.

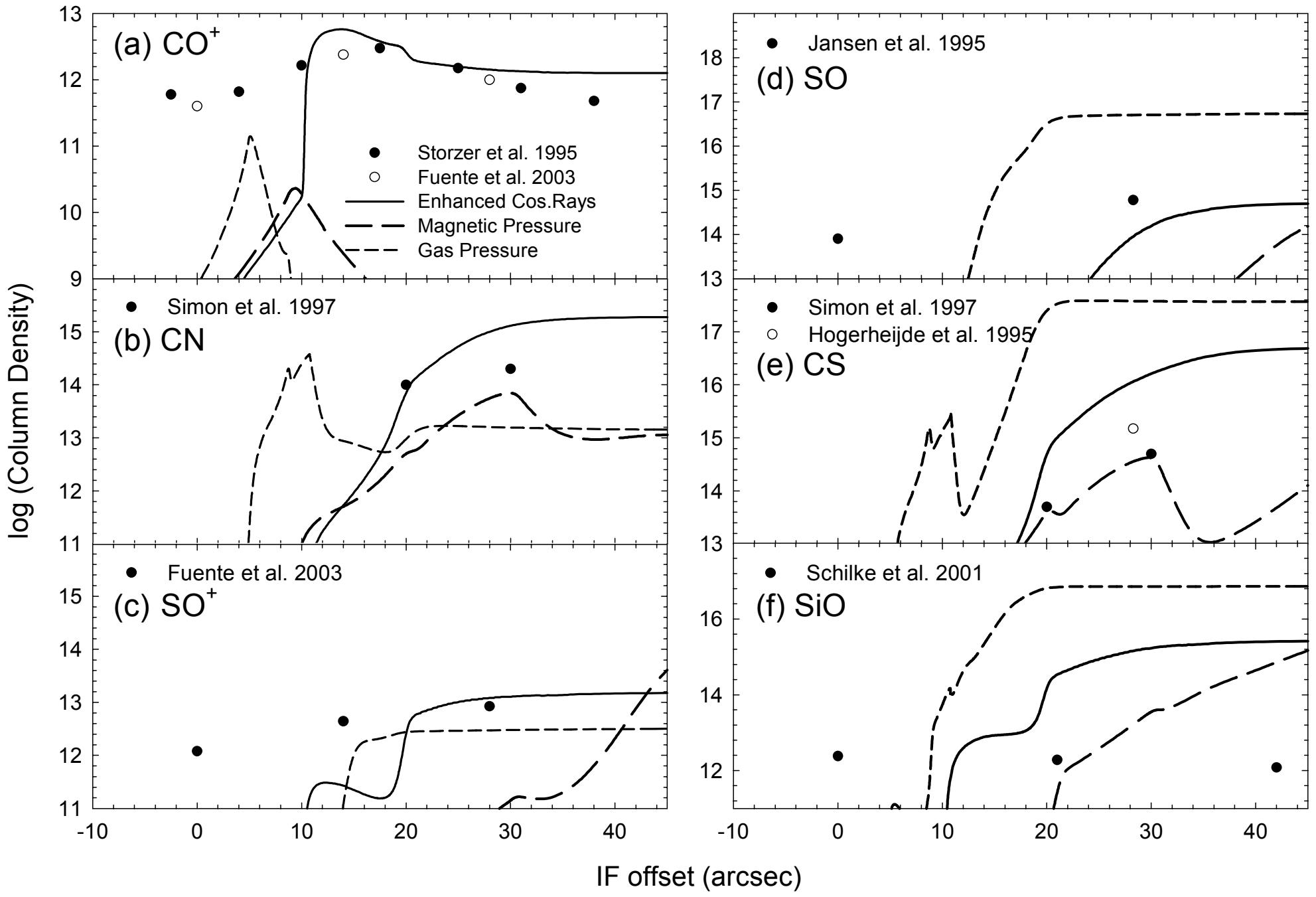

Figure 7. Diatomic molecular column densities (in $cm^{-2}$) for (a) $CO^+$, (b) CN, (c) $SO^+$, (d) SO, (e) CS, and (f) SiO. Modeled and observed values in $cm^{-2}$ as a function of angular projection from the ionization front. Shown are the gas pressure, magnetic pressure and enhanced cosmic ray models.

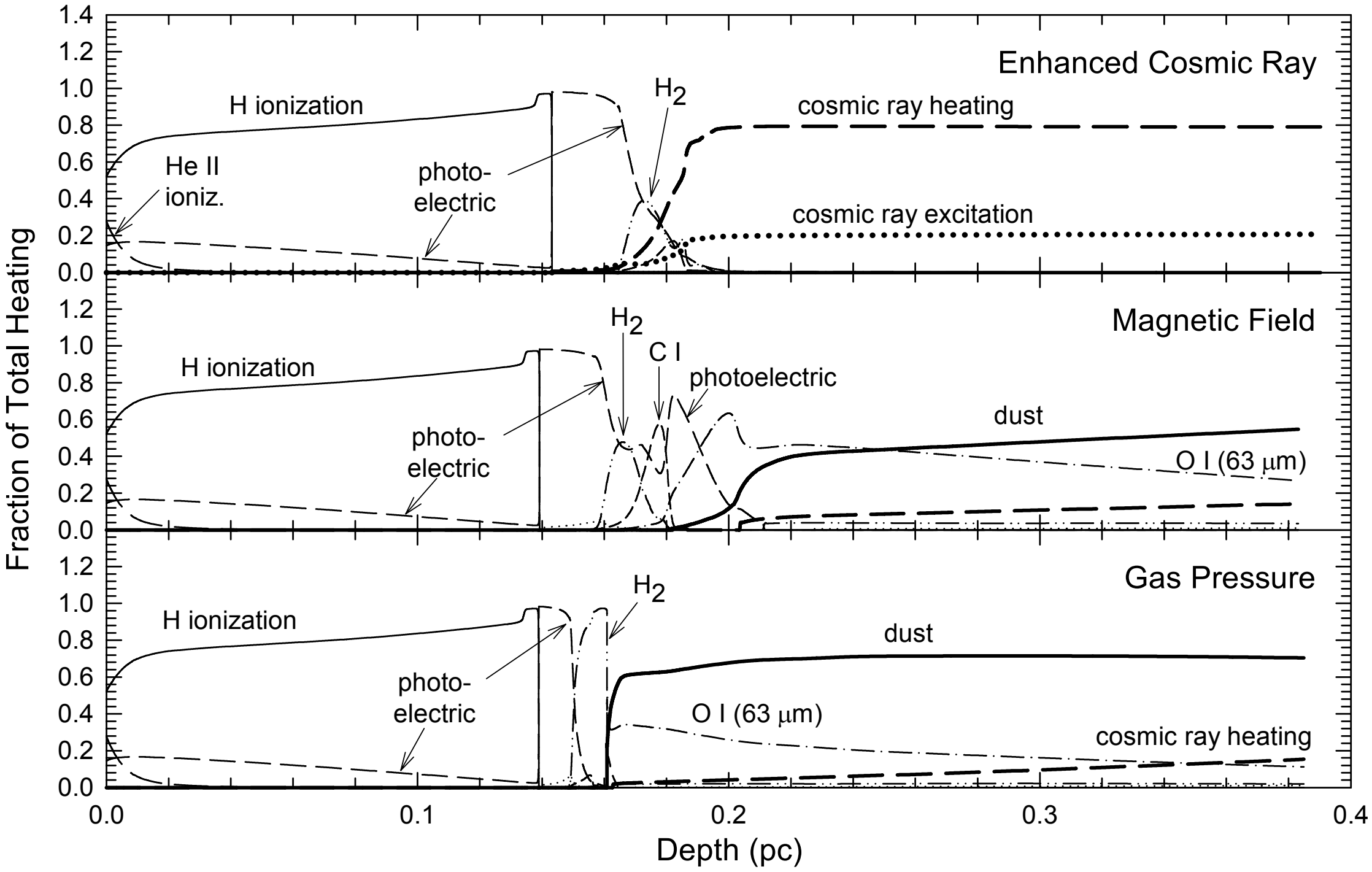

Figure 8. Heating mechanisms in the three models. The line styles indicating each mechanism are the same in each panel. Photoelectric, $H_2$, C I, dust and O I (63 μm) heating are as defined by Tielens & Hollenbach (1985). We also show heating by H I and He II photoionization in the $H^+$ region, and heating of the molecular gas by direct cosmic ray heating and also by cosmic ray excitation of permitted FUV lines.

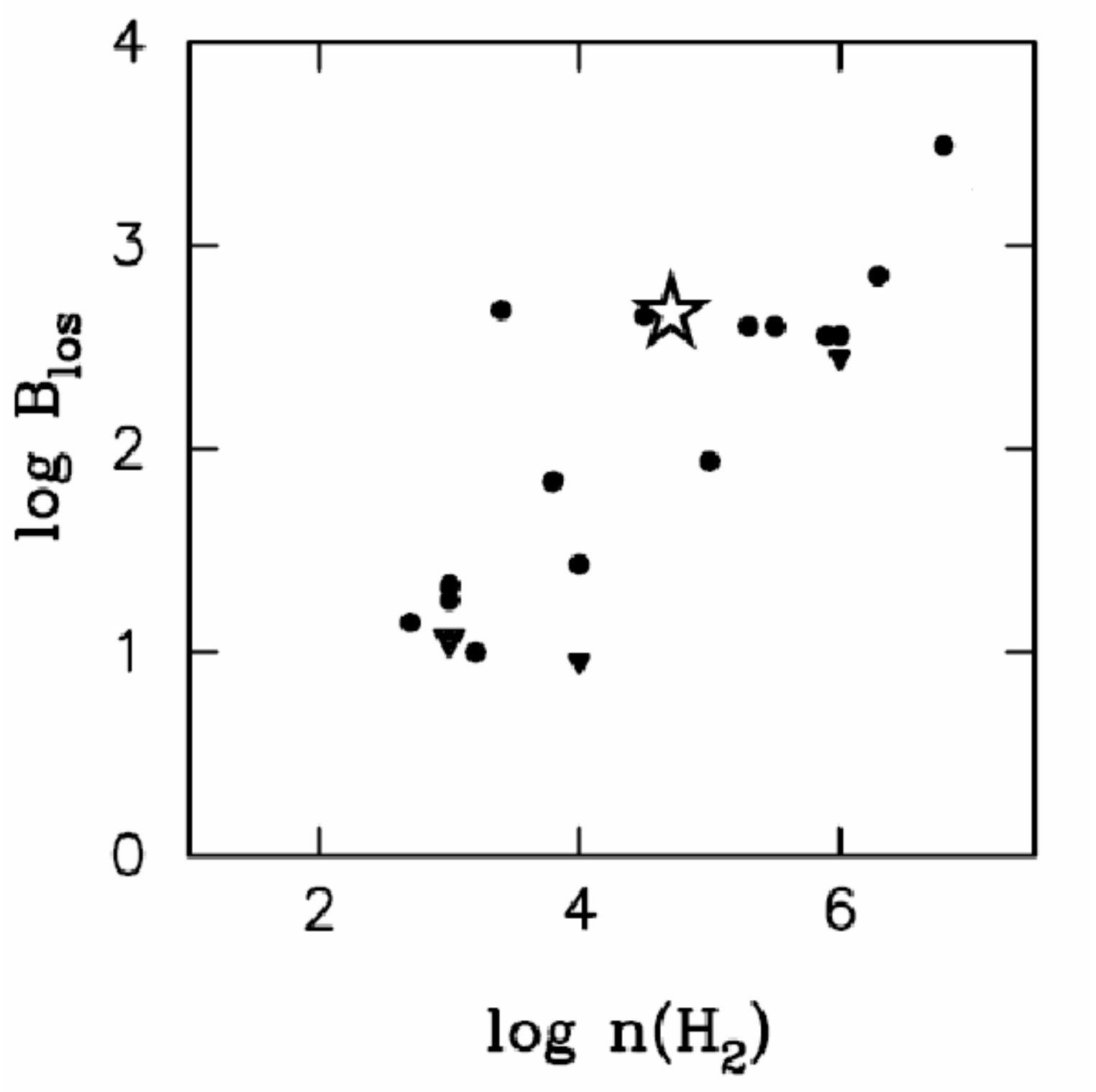

Figure 9. Magnetic field strength vs. $H_2$ gas density, adopted from Crutcher (1999). The star indicates our new result for the Orion Bar. The filled circles are other systems for which $B_{los}$ measurements are available, and the triangles are other systems for which upper limits on $B_{los}$ are available.

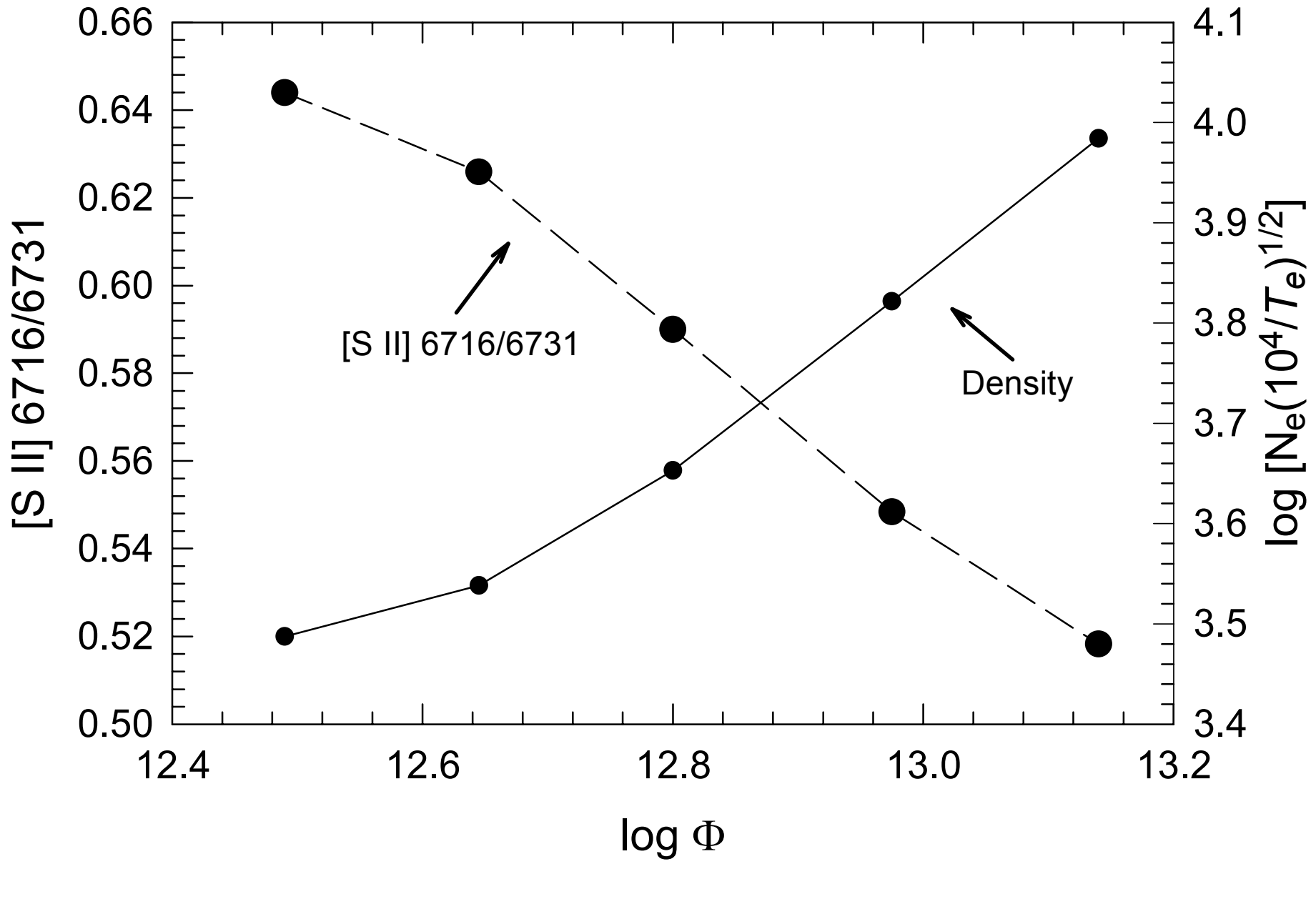

Figure 10. Predicted [S II] ratio and density vs. the incident ionizing photon flux $\Phi$(H).